\documentclass[aps,pr,twocolumn]{revtex4-2}

\usepackage{mathtools}
\usepackage{graphicx}
\usepackage{mathrsfs}
\usepackage{amssymb}
\usepackage{xcolor}
\usepackage{amsmath}
\usepackage{bm}
\usepackage{amsmath}
\usepackage{physics}
\makeatletter\AtBeginDocument{\let\@elt\relax}\makeatother
\usepackage{amsmath}
\usepackage[normalem]{ulem}
\usepackage[unicode=true,colorlinks=true,citecolor=blue,urlcolor=blue]{hyperref}

\def\be {\begin{equation}}
\def\ee {\end{equation}}
\def\bea {\begin{align}}
\def\eea {\end{align}}

\def\bee{\begin{eqnarray}}
\def\eee{\end{eqnarray}}
\def\BC {\begin{cases}}
\def\EC {\end{cases}}

\makeatother

\begin{document}
\title{Magnetophotogalvanic Effects Driven by Terahertz Radiation in CdHgTe Crystals with Kane Fermions }

\author{M. D. Moldavskaya$^1$, L. E. Golub$^1$, V. V. Bel'kov$^2$, S. N. Danilov$^1$,  D. A. Kozlov$^1$, J. Wunderlich$^1$, D.~Weiss$^1$, N.~N.~Mikhailov$^3$, S. A. Dvoretsky$^3$,   S. S. Krishtopenko$^4$, B. Benhamou-Bui$^4$,
F. Teppe$^4$, and S.~D.~Ganichev$^{1}$
}

\affiliation{$^1$Physics Department, University of Regensburg, 93040 Regensburg, Germany}

\affiliation{$^2$Ioffe Institute, 194021 St. Petersburg, Russia}

\affiliation{$^3$Rzhanov Institute of Semiconductor Physics, 630090 Novosibirsk, Russia}

\affiliation{$^4$
Laboratoire Charles Coulomb (L2C), UMR 5221 CNRS-Universit\'{e} de Montpellier, F-34095 Montpellier, France}

\begin{abstract}	
We report on the observation and comprehensive study of the terahertz radiation induced magneto-photogalvanic effect (MPGE) in bulk CdHgTe crystals hosting Kane fermions. The MPGE has been detected in Cd$_{x}$Hg$_{1-x}$Te films with Cd contents $x = 0.15$ and 0.22 subjected to an in-plane magnetic field. At liquid helium temperature we observed multiple resonances in MPGE current upon variation of magnetic field.  In the $x = 0.22$ with noninverted band structure, the resonances are caused by cyclotron resonance (CR) and photoionization of an impurity level.  In the $x = 0.15$ films with an inverted band structure, they originate from the CR and interband optical transitions. Band structure calculated by the Kane model perfectly describes positions of all resonances. In particularly, the resonant MPGE caused by interband transitions excited by THz radiation is caused by the gapless energy spectrum of Kane fermions realized in materials with certain Cd contents and temperature range. In addition to the resonant MPGE current we detected a nonresonant one due to indirect optical transitions (Drude-like). This contribution has a nonmonotonic magnetic field dependence increasing linearly at low magnetic field $B$, approaching a maximum at moderate field and decreasing at high $B$. While the nonresonant MPGE decreases drastically with increasing temperature, it is well measurable up to room temperature. The developed theory demonstrates that the MPGE current arises due to cubic in momentum spin-dependent terms in the scattering probability. The asymmetry caused by these effects results in a pure spin current which is converted into an electric current due to the Zeeman effect.
\end{abstract}

\maketitle
\section{Introduction}
\label{intro}

Three-dimensional topological semimetals are characterized by bulk band crossings in their electronic structures, resulting in gapless electronic excitations and distinctive topological properties that contribute to unusual physical phenomena. The last decade has seen a surge in research activity in this field, driven by accurate theoretical predictions, precise material synthesis, and advanced characterization techniques. Dirac and Weyl materials are notable examples, with pseudo-relativistic band dispersion near the Fermi level, where low-energy fermionic excitations, namely Dirac and Weyl fermions, resemble elementary particles from quantum field theory. Nevertheless, massless 3D Kane fermions, a unique pseudo-relativistic quasiparticle, have been identified in bulk Cd$_{x}$Hg$_{1-x}$Te near the semiconductor-to-semimetal topological phase transition at a critical cadmium concentration of $x_C \approx 0.17$ for a temperature of 2 K, and described by a simplified Kane model~\cite{Orlita2014}. These massless 3D particles feature an energy dispersion with the cones intersected by an additional flat band at the vertex, unlike any known relativistic particle. Unlike Dirac fermions, Kane fermions are not protected by symmetry or topology, but they respect time-reversal symmetry and can be created through appropriate band engineering.

Teppe et al.~\cite{Teppe2016} analyzed the physical properties of these particles, and showed that their rest mass can be inverted near the topological phase transition, while their velocity remains constant, demonstrating their universal relativistic behavior. In bulk CdHgTe alloys, the narrow band-gap (the rest mass of Kane electrons) can indeed be tuned from positive to negative by varying temperature~\cite{Laurenti1990}, chemical composition \cite{Harman1961},  hydrostatic pressure~\cite{Szola2022}, or short-range disorder~\cite{Krishtopenko2022}. Moreover, when the band ordering is inverted, 3D massive Kane electrons are accompanied by 2D topological Volkov-Pankratov states at the interfaces~\cite{Pankratov1987,Krishtopenko2020}  which have also been  observed and characterized in Refs.~\cite{Kazakov2021,Savchenko_2023,Inhofer2017,Otteneder2020a}. Kane fermions are not exclusive to Hg-based materials; they also exist in cadmium arsenide (Cd$_3$As$_2$), coexisting with three-dimensional Dirac fermions when the Fermi level is high in the conduction band~\cite{Akrap2016}. However, their physical properties and theoretical models are still being explored, particularly in CdHgTe compounds. For example, recent theoretical work on the relativistic collapse of their Landau levels under different orientations of crossed electric and magnetic fields suggests that Kane fermions are ``complex'' particle consisting of two nested Dirac particles with different rest masses~\cite{Krishtopenko2022a}.

Experimentally Kane fermions have been studied using a variety of optical and opto-electronic techniques including magneto-spectroscopy through absorption, reflection, or emission experiments, as well as, photoconductivity and photogalvanic effects, see, e.g., Refs.~\cite{Orlita2014,Teppe2016,Galeeva2017,Ruffenach2017, Galeeva2020,Hubmann2020,Otteneder2020a,Szola2022, Dvoretsky2019,Varavin2020}. Moreover,  Kane fermion mobility have been shown to be as high as 10$^6$ cm$^2$/V$\cdot$s near the topological phase transition~\cite{Yavorskiy2018}, resulting in strong nonlinear THz dynamics recently observed in CdHgTe \cite{Soranzio2024}. Furthermore, a giant non-saturating magnetoresistance has been observed in gapless CdHgTe due to the gap opening in Landau-quantized Kane electrons~\cite{Vasileva2020}. Finally, the specific origin of the Kane electrons suppresses non-radiative Auger recombination between their non-equidistant Landau levels, suggesting potential for THz Landau lasers~\cite{But2019}.

In this study, we present the first observation and analysis of magneto-photogalvanic effects (MPGE) in bulk CdHgTe crystals with different cadmium contents. MPGE is an effect where an ac electric field is converted into dc   electric current which requires an external magnetic field, for reviews see, e.g., Refs.~\cite{Ganichev2005,Belkov2008,Ivchenko2018}. Our experimental and theoretical work explores MPGE driven by Drude absorption of THz radiation by Kane fermions and resonant MPGE resulting from inter-Landau level transitions, interband optical transitions, and ionization of impurity levels. The multiple resonances of the MPGE current in the magnetic field have been uniquely identified with the corresponding optical transitions considering the band structure calculated in the frame of $\bm k \cdot \bm p$ theory. The developed theory and experiments show that the MPGE is excited in the bulk of the material and is caused by spin current excitation, which is converted into an electrical current by the Zeeman effect. We show that the pure spin current is caused by the asymmetric scattering of the photoexcited carriers taking into account terms cubic in momentum in the scattering probability.

\begin{figure}
	\centering \includegraphics[width=\linewidth]{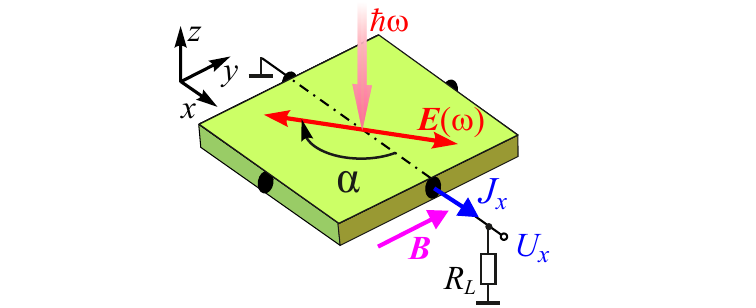}
	\caption{Experimental setup. The sample is excited by normally incident linearly polarized terahertz radiation. In-plane orientation of the radiation electric field $\bm E$ is defined by the azimuth angle $\alpha$ counted from $x$-axis. The magnetic field $\bm B$ is applied in the plane along the $y$-axis. The photosignal is measured  from the pair of contacts made along the $x$-direction as a voltage drop $U_x$ across the load resistance $R_L=50~\Omega$.}
	\label{Fig1}
\end{figure}

\section{Experimental methods and samples}
\label{samples-methods}

MPGE was studied in molecular beam epitaxy grown Cd$_{x}$Hg$_{1-x}$Te films having a common general layer structure and Cd content $x = 0.15$ (sample A) and $0.22$ (sample B). The films were grown on (013)-oriented GaAs substrates followed by a 30~nm ZnTe buffer layer and a 6~$\mu$m layer of CdTe. The film thickness with constant Cd content was 4~$\mu$m (sample A) and 7~$\mu$m (sample B). More details on the sample's composition and parameters can be found in Ref.~\cite{Otteneder2020a} where the same samples were studied.  The square shaped samples with the sides oriented along [$110$] and [$1\bar{1}0$]  have a sample size of $5\times 5$~mm$^2$.  Ohmic indium contacts were soldered to the center of the sample edges, see Fig.~\ref{Fig1}. The transport and magneto-transport characteristics of the samples are presented in the Appendix~\ref{App_A}. The samples were placed in a temperature variable optical cryostat with $z$-cut crystal quartz windows and split coil magnet allowing measurements in the Voigt geometry. In the experiments we used temperatures from 4.2 to 300~K and the magnetic field up to 4~T.

\begin{figure}
	\centering \includegraphics[width=0.8\linewidth]{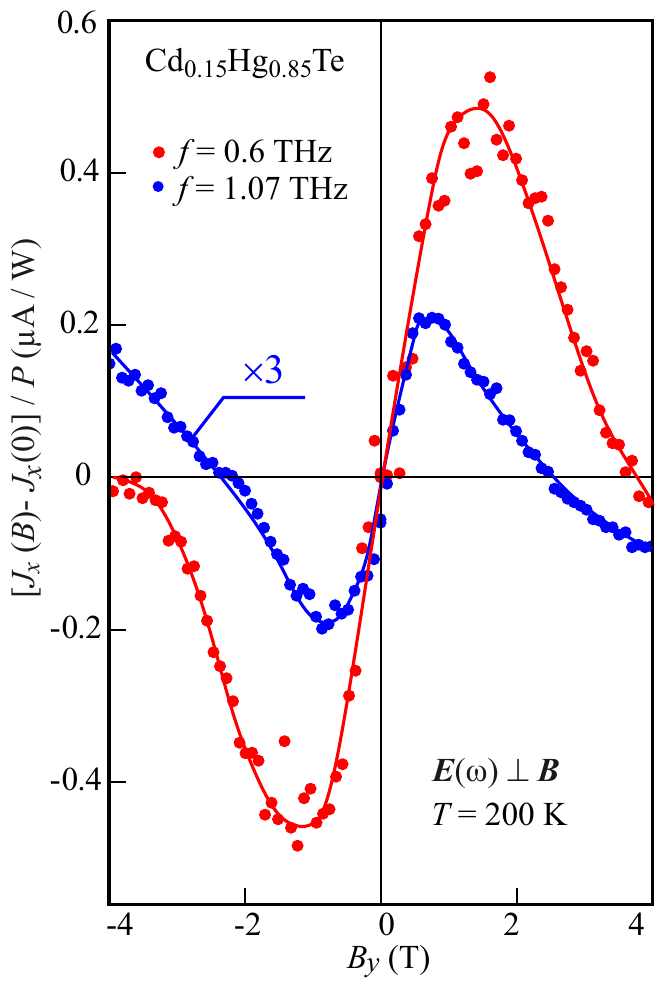}
	\caption{
		Magnetic field dependence of the photocurrent $J_x(B_y)$ measured in the sample A ($x=0.15$) at $T=200$~K. The photocurrent is measured for the radiation electric field perpendicular to the magnetic field direction ($\alpha=0$) and two frequencies $f=0.6$ and 1.07~THz. A small zero magnetic field response $J_x(0)$ is subtracted from the current $J_x(B_y)$ and the data are normalized to the radiation power $P$. Normalized photocurrent at zero magnetic field,  $J_x(0)/P$, and radiation power $P$ in these measurements were  (i) $- 0.1~\mu$A/W and 360 W (red curve, $f=0.6$~THz); (ii) $- 0.014~\mu$A/W and 4200 W (blue curve, $f=1.07$~THz), respectively.  Note that the data for $f=1.07$~THz are multiplied by factor 3.  The main panel shows the photocurrent $J_x$ detected in the direction perpendicular to the magnetic field $B_y$. 
}
	\label{Fig2}
\end{figure}

The photocurrent was studied using a line-tunable  optically pumped pulsed molecular laser with NH$_3$ and CH$_3$F gases as active media~\cite{Ganichev1993,Ganichev1995,Ganichev1998}. The laser operated at $f = 0.6$~THz ($\lambda = 496~\mu$m, $\hbar\omega =2.5$~meV), $f = 1.07$~THz ($\lambda = 280~\mu$m, $\hbar\omega =4.4$~meV), $f = 2.02$~THz ($\lambda = 148~\mu$m, $\hbar\omega =8.4$~meV), and 3.33~THz ($\lambda = 90.5~\mu$m, $\hbar\omega =13.7$~meV). The pump laser was a line-tunable transversely-excited atmospheric pressure (TEA) CO$_2$. The molecular laser generated single pulses with a duration of about 100\,ns and a repetition rate of 1~Hz. The laser radiation power was approximately 100~kW. The peak power of the radiation was monitored by infrared and terahertz photon-drag  detectors~\cite{Ganichev2006,Ganichev1985}. The beam positions and profiles were controlled by a pyroelectric camera. The nearly Gaussian beam  was focused on the sample using an off-axis parabolic mirror resulting in a spot size of 1.5 to 3~mm, depending on the radiation frequency. As we show below, the MPGE effect increases drastically (by four orders of magnitude)  with decreasing temperature and saturates with the radiation intensity $I$ at high $I$. Therefore, to avoid the photocurrent nonlinearity we strongly reduced $I$ for low-$T$ measurements.

\begin{figure}
	\centering \includegraphics[width=\linewidth]{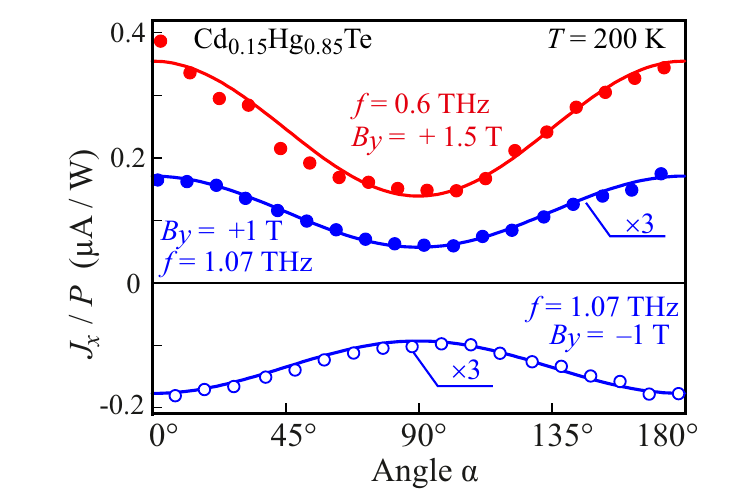}
	\caption{Normalized photocurrent $J_x/P$ as a function of the azimuth angle $\alpha$ measured in sample A ($x=0.15$) subjected to magnetic field $B_y$ and excited by radiation at $f= 0.6$ and 1.07~THz. The data are obtained at $T=200$~K. The radiation power $P$ in these measurements was 360 and 4200 W for the frequencies $f=0.6$ and  $1.07$~THz, respectively. The lines are fits according to Eq.~\eqref{j_perp_alpha}.
	}
	\label{Fig3}
\end{figure}

\begin{figure}
	\centering \includegraphics[width=\linewidth]{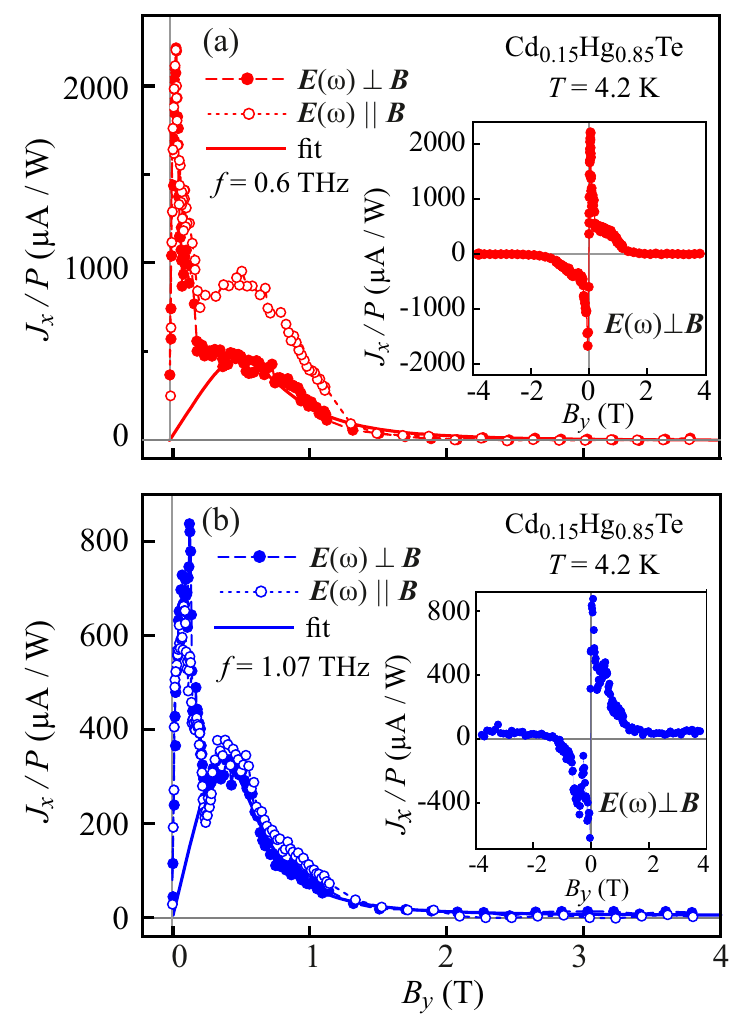}
	\caption{Magnetic field dependencies of the photocurrent $J_x$ normalized to the power $P$ obtained at $T=4.2$~K and for  the radiation electric field perpendicular and parallel to the magnetic field $B_y$.   Data are shown for the radiation frequencies $f=0.6$, panel (a), and 1.07~THz, panel (b). The insets show $J_x(B_y)$ for both magnetic field polarities  and $\bm E(\omega) \perp \bm B$. The radiation power $P$ in these measurements were 0.06 W for both frequencies. The solid curves are fits according to Eq.~\eqref{empiric}.
	}
	\label{Fig4}
\end{figure}

The experiments were performed using normal incident  linearly polarized radiation, see Fig.~\ref{Fig1}. In order to vary the angle between the polarization vector of the linearly polarized light and the magnetic field, we used crystalline quartz $\lambda/2$-plates. Rotation of the half-wave plate allowed us to vary the azimuth angle $\alpha$ from $0^\circ$ to $180^\circ$, covering all possible orientations of the electric field vector in the ($xy$) plane with respect to the in-plane magnetic field. In the following we choose $y \parallel \bm B$ and the angle $\alpha = 0$  is zero along the $x$-direction,   perpendicular to vector $\bm B$, see Fig.~\ref{Fig1}. The terahertz radiation induces free carrier absorption in the lowest conduction subband $e1$ because the photon energies are smaller than the energy gap. For certain magnetic field strengths, however, the radiation also excites inter-Landau level transitions, causing the resonant MPGE current. The photocurrent $J_x$  was measured in unbiased structures from the voltage drop $U$ across a load resistor $R_L = 50~\Omega$ in closed circuit configuration. The voltage was recorded with a storage oscilloscope and the photocurrent was calculated according to   $J_x = U/R_\parallel$, where $R_\parallel = R_s\times R_L / (R_s+ R_l)$ and $R_s$ is the sample resistance. The measured photocurrent pulses of 100~ns duration reflected the corresponding laser pulses.

\section{Experiment}
\label{Experiment}

\subsection{Results for Cd$_{x}$Hg$_{1-x}$Te films with $x=0.15$}

We begin with the data obtained on sample A with $x = 0.15$. Figure~\ref{Fig2} shows the magnetic field dependence of the photocurrent excited in the direction perpendicular to the in-plane magnetic field $B_y$ by linearly polarized radiation with $\bm E(\omega) \perp \bm B$. The data are obtained at $T=200$~K. To emphasize the magnetic field induced photoresponse in this plot we have subtracted the photocurrent generated at $B=0$. The photocurrent is odd in a magnetic field and depends non-monotonically on $B$: at low magnetic fields it increases almost linearly, approaches a maximum, and then decreases and changes its sign (see the data for $f=1.07$~THz). Such a qualitative behavior is observed for all frequencies studied.  We observed that increasing the frequency reduces the photocurrent  magnitude at the maximum, and shifts the maximum as well as the inversion point to lower magnetic fields. At low power, the photocurrent depends linearly on the radiation power, and saturates as the power increases. To study the photoresponse in the linear regime, in all measurements below we first measured the power dependences and then substantially reduced the radiation power until reaching the regime $J \propto P \propto E^2(\omega)$~\footnote{At low temperatures we were forced to reduce power by $10^6$. The observed nonlinearity is of independent interest but is out of scope of the present paper.}.

The observed magnetic field induced photocurrent consists of a polarization independent part and  a part varying with the azimuth angle $\alpha$ as $\cos (2 \alpha)$. Figure~\ref{Fig3} shows, as an example, several characteristic traces obtained at two frequencies and for the magnetic field strength corresponding to the signal maximum. The data obtained at $B_y = \pm 1$~T and $f=1.07$~THz demonstrate that both, polarization dependent and independent photocurrents, are odd in $B_y$.

Cooling to 4.2 K results in a drastic increase of the photocurrent amplitude by more than three orders of magnitude, compare  Figs.~\ref{Fig4} and~\ref{Fig2}. Moreover,  the nonmonotonic photocurrent detected at high temperature is now accompanied by resonances  that appear at low magnetic fields and have about twice the magnitude of the nonresonant one. The insets in Fig.~\ref{Fig4} show that both types of the photocurrent reverse sign when the  magnetic field  polarity is changed. 

Comparing Figs.~\ref{Fig2} and~\ref{Fig4}, we see that lowering  temperature causes  the maximum of the nonresonant photocurrent narrow and to shift to lower magnetic fields.  Furthermore, at high field, the nonresonant current disappears  and the sign reversal with increasing magnetic field strength observed at high temperatures, is absent. 

\begin{figure}[t]
	\centering \includegraphics[width=\linewidth]{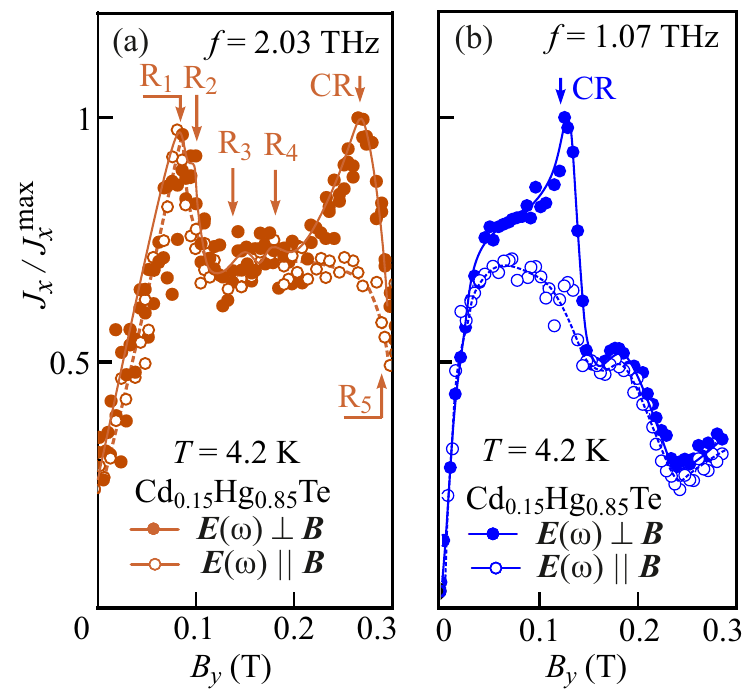}
	\caption{Low-field zoom of the magnetic field dependencies of the photocurrent $J_x$ normalized to the power $P$ obtained  in sample A at $T=4.2$~K. The data are presented for the radiation electric fields perpendicular and parallel to the magnetic field $\bm B$ for the radiation frequencies $f=2.03$, panel (a), and 1.07~THz, panel (b). Downward arrows indicate the positions of the CR- and R-type resonances, for the corresponding magnetic fields see Table~\ref{table1}. The low magnetic field data with higher resolution (smaller magnetic field step is used) are shown in Fig.~\ref{Fig_JB015} and the corresponding optical transitions are sketched in the calculated Landau level diagrams and discussed in the Sec.~\ref{discussion}. The radiation power $P$ in these measurements were 0.1 and 10 W for the frequencies $f=1.07$ and  $2.03$~THz, respectively.  
	}
	\label{Fig5}
\end{figure}

\begin{figure}[h]
	\centering \includegraphics[width=\linewidth]{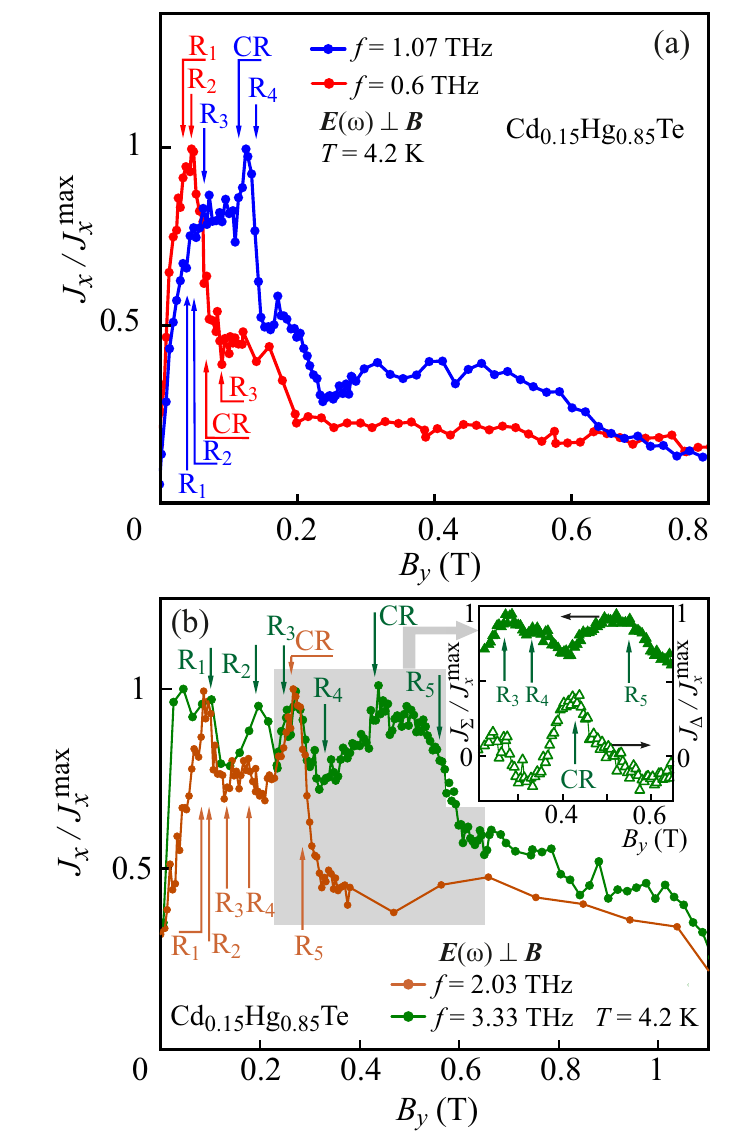}
	\caption{ Magnetic field dependencies of the photocurrents at frequencies $f=1.07$ and 0.6~THz (a), and $f=2.03$ and 3.33~THz (b) measured at $T=4.2$~K with polarization $\bm E(\omega) \perp \bm B$. The photocurrent $J_x$ is normalized by the  maximum photocurrent value $J_x^{\rm max}$.  Downward arrows indicate the positions of the CR- and R-type resonances, for the corresponding magnetic fields see Table~\ref{table1}. Corresponding optical transitions are sketched in the calculated Landau level diagrams and discussed in the Sec.~\ref{discussion}. The inset shows the polarization independent and the polarization dependent photocurrents calculated according to $J_\Sigma = [J(\alpha = 0) + J(\alpha = 90^\circ)]/2$ and $J_\Delta = [J(\alpha = 0) - J(\alpha = 90^\circ)]/2$. The data are obtained at a higher resolution (smaller magnetic field step is used), and shown for the  magnetic field range highlighted by the gray area where CR is expected. The plot is used to resolve the CR resonance, which vanish for $\bm E(\omega) \parallel \bm B$ ($\alpha = 90^\circ$), hidden by the neighboring polarization independent $\rm R_4$- and $\rm R_5$-resonances. 
}
	\label{Fig_JB015}
\end{figure}

Figure~\ref{Fig5} shows a zoom of the low magnetic field range where the photocurrent resonances are detected. Experiments are performed for two different orientations of the radiation electric field with respect to the magnetic field ($\bm E\parallel \bm B$ and $\bm E \perp \bm B$). The data reveal  two types of resonances on top of the nonresonant photoresponse: (i) independent of the radiation polarization (we will label them $R$), and (ii) excited only for $\bm E \perp \bm B$  (labeled as CR). As we show in Sec.~\ref{discussion}, CR-type resonances arise  from transitions between the electron Landau levels (cyclotron resonances), while R-type resonances are caused by interband transitions between levels of heavy holes and electrons in an inverted band structure. Figure~\ref{Fig_JB015}  shows the photocurrents measured with higher resolution in the low magnetic field region. The data are   shown for four frequencies at CR-active polarization. It can be seen that with increasing frequency, more R-type resonances appear on the $J_x(B_y)$ dependence.  CR-type resonances are clearly seen at 1.07 and 2.03 THz. At a frequency of  3.33~THz, the CR-type resonance is visible as a difference between the active and inactive polarization, see the inset in Fig.~\ref{Fig_JB015}(b).

\subsection{Results for Cd$_{x}$Hg$_{1-x}$Te films with $x=0.22$}

\begin{figure}
	\centering \includegraphics[width=\linewidth]{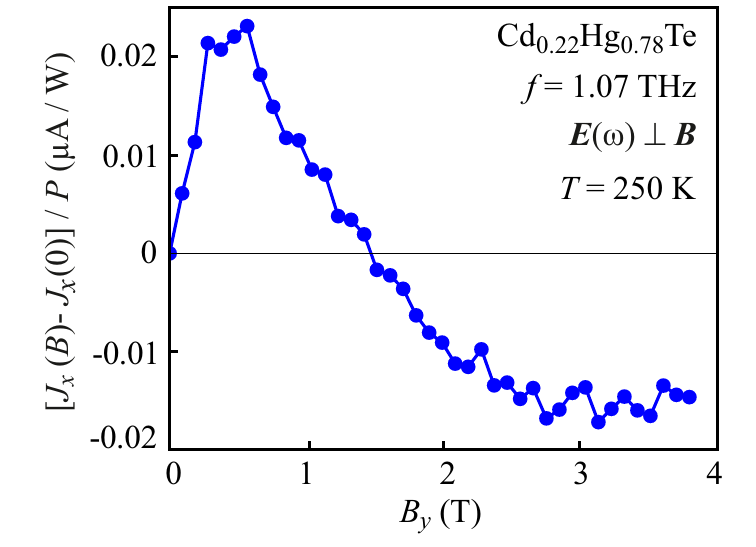}
		\caption{
			Magnetic field dependence of the normalized photocurrent $[J_x(B_y) - J_x(0)]/P$ measured for sample B ($x=0.22$), ${\bm E}(\omega) \perp {\bm B}$, frequency $f=0.6$~THz,  and power $P=0.6$~W. 
	}
	\label{Fig10}
\end{figure}

\begin{figure}
	\centering \includegraphics[width=\linewidth]{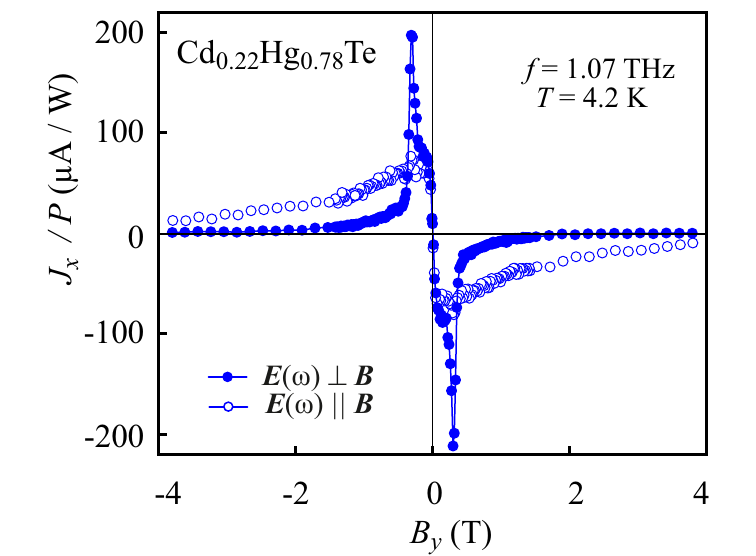}
	\caption{Magnetic field dependence of the normalized photocurrents $J_x(B_y)/P$ measured in the sample B ($x=0.22$) at $T=4.2$~K. The data are presented for the radiation electric field $\bm E(\omega)$ perpendicular and parallel to the magnetic field $\bm B$ for the radiation frequency $f=1.07$~THz and power $P =  0.6$~W.
	}
	\label{Fig8}
\end{figure}

\begin{figure}[t]
	\centering \includegraphics[width=\linewidth]{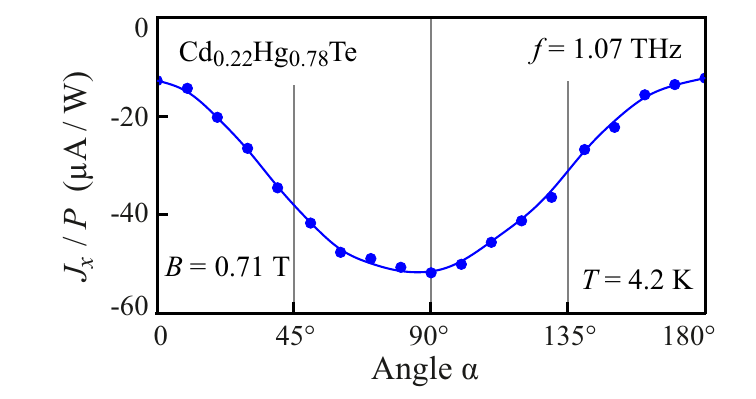}
	\caption{Dependence of the normalized photocurrent $J_x/P$ on the azimuth angle $\alpha$ measured in sample B ($x=0.22$) subjected to a magnetic field $B_y=0.71$~T and excited by radiation with $f= 1.07$~THz. The data were obtained for $T=4.2$~K.  The radiation power $P$ in this measurement was 0.6~W. Lines are fits to Eq.~\eqref{j_perp_alpha}.
	}
	\label{Fig11}
\end{figure}

Now we turn to the results obtained for sample B with $x=0.22$. Figure~\ref{Fig10} shows the magnetic field dependence of the photocurrent detected at $250$~K. As for sample A ($x=0.15$), the signal has a nonmonotonic magnetic field dependence at high temperatures and changes its sign at high magnetic fields. Reducing the temperature down to $T=4.2$~K strongly increases the magnitude of the nonresonant photocurrent maximum, and, for $\bm E \perp \bm B$ a nonresonant photocurrent is superimposed on the resonant one, see Fig.~\ref{Fig8}. Note that at low temperatures the currents detected in sample B have the opposite sign to those observed at $T =250$~K and to the  photocurrents generated in sample A. The signal shows a characteristic polarization dependence of the photocurrent. The signal is odd in the magnetic field, and  its polarization dependence is well described by the function $J\propto \cos (2\alpha)$, see Figs.~\ref{Fig8} and~\ref{Fig11}. 

\begin{figure}
	\centering \includegraphics[width=\linewidth]{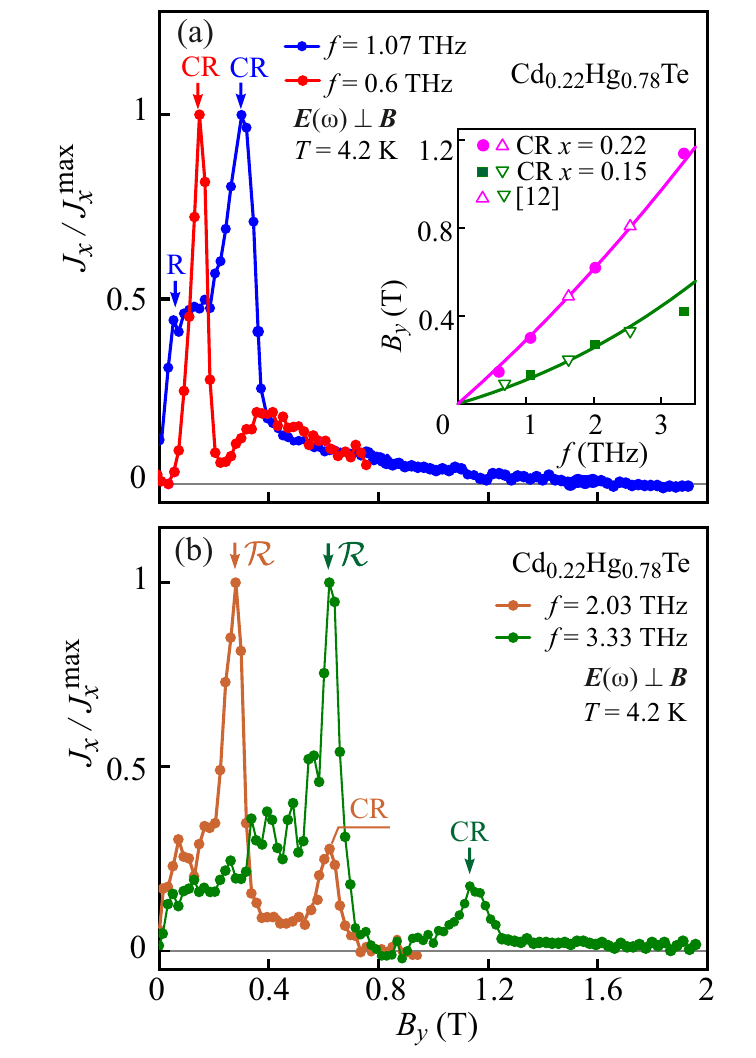}
	\caption{   Low-field part of the magnetic field dependencies of the photocurrents $J_x(B_y)$ obtained at $T=4.2$~K and normalized to their maxima. The data are presented for sample B ($x=0.22$), the radiation electric field perpendicular  to the magnetic field $\bm B$, and  for the radiation frequencies $f=0.6$, 1.07, 2.03, and 3.33~THz. The radiation power $P$ in these measurements was  0.065 W ($f=0.6$~THz); 0.6 W ($f=1.07$~THz); 0.02~W ($f=2.02$~THz); and 0.06~W ($f=3.3$~THz).  	Arrows labeled CR show resonances, that are absent for the radiation electric field parallel  to the magnetic field $B_y$. Arrows labeled  by $\cal R$  show resonances, that are insensitive to the radiation polarization. 
 The corresponding magnetic fields are given in Table~\ref{table2}.	The inset shows the frequency dependencies of the CR position. Here open triangles are the CR positions obtained in Ref.~\cite{Otteneder2020a} in the radiation transmission experiments.  Lines are calculated frequency dependencies of the CR  positions.
	}
	\label{Fig9}
\end{figure}

As mentioned above, in addition to the non-resonant photocurrent, we also detected a resonant contribution in sample B at $T=4.2$~K.
The resonant peak is observed at CR-active polarization $\bm E(\omega) \perp \bm B$ and is absent at passive polarization $\bm E(\omega) \parallel \bm B$, see Fig.~\ref{Fig8}. The frequency dependence of the CR resonance positions for both samples is shown in the inset of Fig.~\ref{Fig9}\,(a), demonstrating that the resonance magnetic fields increase with increasing frequency. Polarization-independent resonances denoted as $\cal R$ are also observed in the sample B at high frequencies, see Fig.~\ref{Fig9}.  However, unlike sample A where multiple R-type of resonances were detected, see Fig.~\ref{Fig_JB015}, the resonances  $\cal R$ in sample B appear as only a single line for each frequency. Note that the maximum photocurrent value at the same frequency in the sample B is 4-5 times smaller than in the sample A.

\section{Theory}
\label{theory}

The MPGE current can arise owing to two distinct microscopic grounds. The first mechanism is related to spin properties of carriers. With account for the spin-orbit interaction, the fluxes of particles in each spin subband, $\bm i_\uparrow$ and $\bm i_\downarrow$, are generated under optical absorption, and the pure spin current 
\begin{equation}
	\label{J_def}
	\bm{\mathcal J}={\bm i_\uparrow- \bm i_\downarrow\over 2}
\end{equation}
is formed~\cite{Ganichev2006}.
In bulk CdHgTe, the 
spin fluxes appear under radiation absorption due to 
the odd in momentum terms in the probability of the electron-phonon interaction. In the crystals of $T_d$ symmetry the corresponding scattering matrix element has the following form
\begin{equation}
	\label{U}
	U_{\bm k' \bm k}= U_0\qty[1+\xi\sum_{i=x',y',z'}\sigma_{i}K_i \qty(K_{i+1}^2-K_{i+2}^2)],
\end{equation}
where $U_0(\bm k' - \bm k)$ is the electron-phonon scattering matrix element without account for the non-centrosymmetry of the crystal,
$x' \parallel [100]$, $y' \parallel [010]$, $z' \parallel [001]$ are the cubic axes,
$\sigma_{i}$ are the Pauli matrices, cyclic permutation of the indices is assumed, $\xi$ is a material constant, and
\begin{equation}
	\label{K_sum}
	\bm K=\bm k + \bm k'.
\end{equation}
It follows from this expression that
electrons with spins along and opposite to $\bm B \parallel y$ have different scattering 
matrix elements $U_{\bm k' \bm k}^\pm$ given by
\begin{equation}
	\label{Upm}
	U_{\bm k' \bm k}^\pm= U_0\qty(1\pm \xi \sum_{i,j,l=x,y,z}a_{ijl}K_iK_jK_l),
\end{equation}
where
the coefficients $a_{ijl}$ are determined by 
the orientation of the coordinate system $(x,y,z)$ in respect to the cubic axes.

\begin{figure}[t]
	\centering \includegraphics[width=\linewidth]{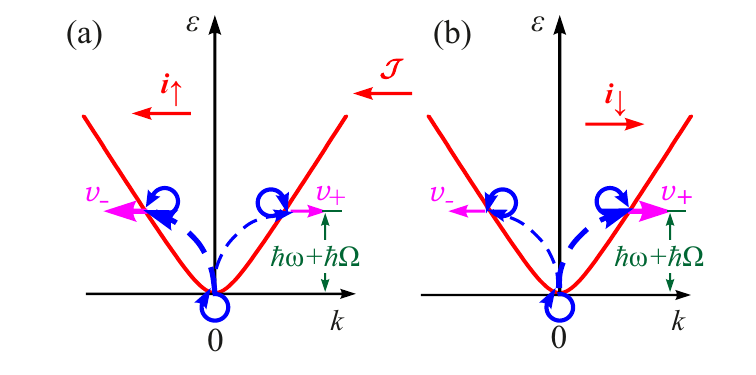} 
	\caption{  Model of a pure spin photocurrent  at zero magnetic field. 	Solid red curves show the Kane dispersion. Curved dotted and solid blue arrows sketch, respectively, scattering on phonons with the energy $\hbar \Omega$, and optical transition involving different virtual states in the same subband. Different thickness indicates difference in the probabilities of the corresponding processes.  Horizontal magenta arrows show the resulting velocities for positive, $v_+$, and negative, $v_-$, wavevectors  $k$. Red horizontal arrows show spin fluxes generated in the spin-up,  $\bm i_\uparrow$, and spin-down, $\bm i_\downarrow$, subbands as well as the total spin current $\bm{\mathcal J}$.
	}
	\label{Fig_Zero_field_Scheme}
\end{figure}

We develop a theory of MPGE for the relevant to the experiment case of Drude-like  intraband radiation absorption.
It is important that this type of absorption is caused not only by the carrier-photon interaction, but also by scattering. The cubic in momentum terms of opposite sign in the electron-phonon scattering matrix elements~\eqref{Upm} 
result in an enhanced probability of the optical transition with the final state at positive (negative) momenta in the spin-down (spin-up) subband. 
Consequently, occupation of the final states is different, Fig.~\ref{Fig_Zero_field_Scheme}.
This imbalance in each subband results in the oppositely directed spin fluxes $\bm i_\uparrow$ and $\bm i_\downarrow$, and in the pure spin current $\bm{\mathcal J}$, Eq.~\eqref{J_def}.

In the absence of magnetic field, the charge current 
\begin{equation}
	\label{j_charge}
	\bm j=e(\bm i_\uparrow + \bm i_\downarrow)
\end{equation}
is not generated due to compensation of the fluxes of particles with opposite spins.
The MPGE current is produced from the pure spin current by the Zeeman effect.
Application of a magnetic field $\bm B$ leads to the net carrier spin polarization $S$.
For degenerate three-dimensional carriers  we have
\begin{equation}
	\label{S_deg}
	S = -{1-\sqrt{1-g\mu_{\rm B}B/\varepsilon_{\rm F}} \over 2(1+\sqrt{1-g\mu_{\rm B}B/\varepsilon_{\rm F}})},
\end{equation}
where $\varepsilon_{\rm F}$ is the Fermi energy and $g$ is the electron Land\'e factor. Here we assumed $\abs{g\mu_{\rm B}B}<\varepsilon_{\rm F}$, otherwise $S=-1/2$. At non-degenerate statistics, the average spin reads
\begin{equation}
	\label{S_nondeg}
	S = -{1\over 2} \tanh({g\mu_{\rm B}B \over 2 k_{\rm B}T}).
\end{equation}
Different numbers of the spin-up and spin-down carriers mean that the spin fluxes $\bm i_\uparrow$, $\bm i_\downarrow$ have different absolute values. As a result, they do not compensate each other any more, and the charge current is generated under radiation absorption. The electric current density 
is given by a product of the pure spin current $\bm{\mathcal J}$ and the average spin $S$~\cite{Ganichev2006}:
\begin{equation}
	\label{j_spin_current}
	\bm j = 4e S \bm{\mathcal J}.
\end{equation}

\begin{figure}[t]
	\centering \includegraphics[width=\linewidth]{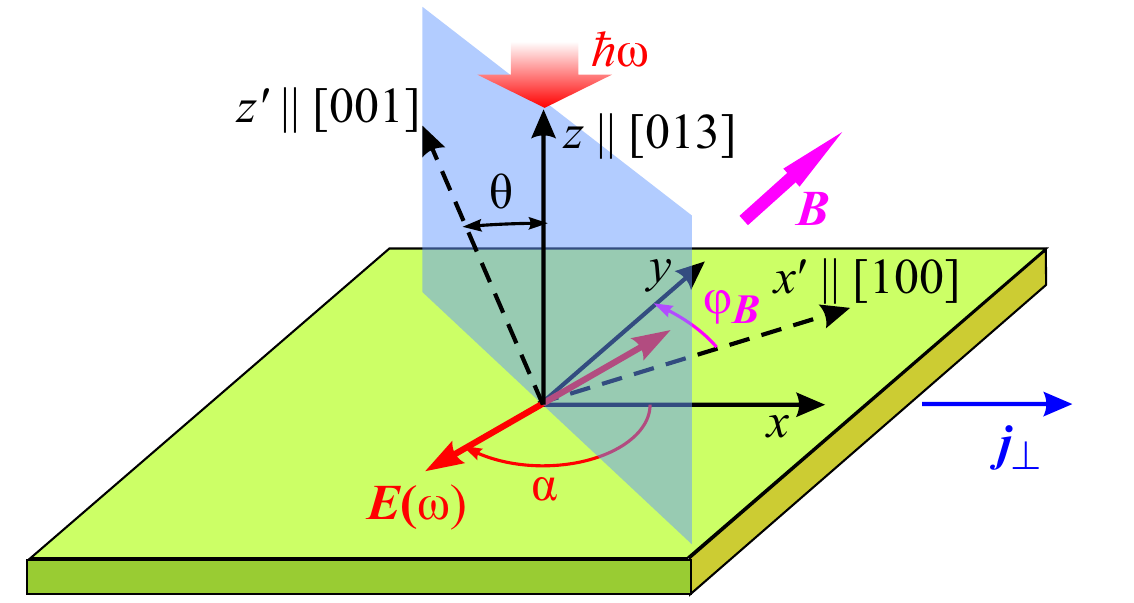}
	\caption{Definition of the coordinate axes. In the Voigt geometry with radiation normally incident on the (013) sample surface, we introduce  the axes $(x,y)$ in the sample plane with $y \parallel \bm B$. The angle $\varphi_{\bm B}$ is an angle between $\bm B$ and the axis  $x' \parallel [100]$. The axes $z \parallel [013]$ and $z' \parallel [001]$ lie in the plane perpendicular to the axis $x'$ (blue), and the angle between them $\theta = \arctan(1/3)\approx 18.4^\circ$. The angle $\alpha$ defines the polarization direction with $\alpha=0$ corresponding to $\bm E \perp \bm B$.
	}
	\label{Fig14}
\end{figure}

Competing is MPGE's orbital mechanism. In the magnetic field, there are spin-independent corrections to the scattering rates of the carriers, which are odd in $\bm B$ and in the carriers' momentum.
Such a diamagnetic contribution to the scattering matrix element  in the crystals of $T_d$ symmetry
is given by
\begin{equation}
	\label{energy_dia}
	U_{\bm k' \bm k}= U_0\qty[1+\zeta\sum_{i=x',y',z'}B_{i}K_i \qty(K_{i+1}^2-K_{i+2}^2)],
\end{equation}
where $\bm K$ is given by Eq.~\eqref{K_sum}, and $\zeta$ is a material constant.
Accordingly, the probability of radiation absorption is different for positive and negative carriers' momenta, and an asymmetric distribution of the carriers is formed in the momentum space as a result of optical absorption. This means an appearance of a net charge current in both spin subbands.

Both mechanisms of MPGE give comparable contributions to the charge current. In order to estimate the MPGE current we consider below the spin mechanism. 
We consider the limits of weak and strong magnetic fields and derive the frequency, magnetic field and radiation polarization dependencies of the MPGE current.

In the experiment, there are resonant and nonresonant contributions to the MPGE current. The nonresonant part is present at any polarization but it is polarization-dependent. The resonant contributions are of two types: 
one (R) exists at any polarization, and another (CR) is present only if the radiation electric field is not parallel to $\bm B$ in the Voigt geometry. 
Below we develop a theory of MPGE for Kane fermions and consider 
the nonresonant part of the MPGE current caused by Drude-like intraband absorption in the conduction band.
Then we investigate the CR and R peaks in the photocurrent caused by direct optical transitions between the conduction Landau subbands and interband transitions from the heavy-hole to the conduction band, respectively.

\subsection{Classical magnetic fields}

CdHgTe crystals have $T_d$ point symmetry. This means that the MPGE current density, in the linear in the magnetic field regime, is described by two linearly independent constants, $\Phi_1$ and $\Phi_2$~\cite{Ivchenko1988}
\begin{align}
	\label{phenom_cubic}
	j_{x'} & = \Phi_1  e_{x'}(B_{y'}e_{y'}-B_{z'}e_{z'})I+ \Phi_2 B_{x'}(e_{y'}^2-e_{z'}^2)I,\nonumber\\
	j_{y'} & = \Phi_1 e_{y'}(B_{z'}e_{z'}-B_{x'}e_{x'})I+ \Phi_2 B_{y'}(e_{z'}^2-e_{x'}^2)I, \nonumber\\
	j_{z'} & = \Phi_1 e_{z'}(B_{x'}e_{x'}-B_{y'}e_{y'})I+ \Phi_2 B_{z'}(e_{x'}^2-e_{y'}^2)I.
\end{align}
Here $x',y',z'$ are the cubic axes, see Eq.~\eqref{U}, $\bm e$ is the polarization vector, and $I$ is the radiation intensity.
This expression and the phenomenology for the Linear photogalvanic effect (LPGE) at $B=0$ show that $\Phi_1$ has a contribution from the Hall component of the LPGE current while $\Phi_2$ describes the new contribution generated by the magnetic field.

Experiments are performed in the Voigt geometry with radiation normally incident on the (013) sample surface
Therefore we introduce the axes $(x,y)$ in the sample plane with $y \parallel \bm B$.
For the MPGE current   perpendicular to $\bm B$ we have from Eqs.~\eqref{phenom_cubic}
\begin{multline}
	\label{j_perp}
	j_x = -{B\cos{2\theta}\over 2} I\bigl\{ \Phi_2\sin{2\varphi_{\bm B}}
	\\ +
	\Phi_1 \qty[2e_xe_y\cos{2\varphi_{\bm B}}-(e_x^2-e_y^2)\sin{2\varphi_{\bm B}}]\bigr\},
\end{multline}
where $\varphi_{\bm B}$ is an angle between $\bm B$ and the  $x' \parallel [100]$ axis,
and $\theta = \arctan(1/3)\approx 18.4^\circ$ is the angle between $z \parallel [013]$ and $z' \parallel [001]$, see Fig.~\ref{Fig14}.

It follows from Eqs.~\eqref{S_deg} and~\eqref{S_nondeg} that, in the linear in the magnetic field regime, the spin is given by
\begin{equation}
	\label{S_low_field}
	S = -{g\mu_{\rm B}B\over 4\tilde{\varepsilon}},
\end{equation}
where $\tilde{\varepsilon}$ equals to $2\varepsilon_{\rm F}$ and $k_{\rm B}T$ at degenerate and non-degenerate statistics, respectively. Therefore in low fields, when the electric current is linear in $B$, the pure spin current $\bm{\mathcal J}$ is to be calculated at $B=0$, see Eq.~\eqref{j_spin_current}. 
The zero-field 
fluxes of particles in the spin subbands are given by~\cite{Belkov2005}
\begin{equation}
	\label{J}
	\bm i_s = \sum_{\bm k, \bm k'} W^{(s)}_{\bm k'\bm k}
	[\bm v_{\bm k'}\tau(\varepsilon_{k'})-\bm v_{\bm k}\tau(\varepsilon_{k})],
\end{equation}
where summation is performed over wavectors of electrons in the $s$'th spin subband, $\varepsilon_{k}$ is the Kane dispersion in the conduction band, Fig.~\ref{Fig13}
\begin{equation}
	\varepsilon_k=\varepsilon_g/2+\sqrt{(\varepsilon_g/2)^2+(\hbar v_0 k)^2}
\end{equation}
with $\varepsilon_g$ and $v_0$ being the energy gap and Kane velocity,
the electron velocity 
$\bm v_{\bm k}=\hbar v_0^2 \bm k/(\varepsilon_k-\varepsilon_g/2)$, and
$\tau$ is the energy dependent momentum relaxation time. The probability of the phonon-scattering assisted intra-band optical transition reads~\cite{Belkov2005}
\begin{multline}
	\label{W}
	W^{(s)}_{\bm k'\bm k} = {2\pi\over \hbar} \sum_{\nu = \pm} 
	\qty[f_k(1-f_{k'})N_\nu-f_{k'}(1-f_k)N_{-\nu}] 
	\\
	\times \abs{{eU^s_{\bm k'\bm k}\over \hbar \omega^2}\bm E \cdot (\bm v_{\bm k}-\bm v_{\bm k'})}^2  \delta(\varepsilon_{k}+\hbar\omega -\nu\hbar\Omega - \varepsilon_{k'}),
\end{multline}
where  $U^s_{\bm k'\bm k}$ is the scattering matrix element Eq.~\eqref{Upm},
$N_\nu=N+(1+\nu)/2$, $N$ and $\Omega$ are the occupation number and frequency of phonons with the wavevector $\bm q = \bm k'-\bm k$, $f_k$ is the Fermi-Dirac distribution function,  and $\omega$ is the radiation frequency.

Treating the $\xi$-linear odd in momentum terms in the scattering matrix element~\eqref{Upm} as small perturbations, we calculate the spin fluxes~\eqref{J} and then the pure spin current~\eqref{J_def}.
In order to find $\Phi_2$, we calculated $\mathcal J_{z'}$ for the electrons with spins along $z'$ axis and $\bm e \parallel x'$. To obtain $\Phi_1$, we assumed $e_{x'}, e_{z'} \neq 0$ and calculated $\mathcal J_{x'}$.
Then, taking $S$ in the form of Eq.~\eqref{S_low_field}, we get the electric current from Eq.~\eqref{j_spin_current}.
As a result, we obtain $j_{z'}=\Phi_2 B_{z'}e_{x'}^2I$ and $j_{x'}=-\Phi_1 B_{z'}e_{x'}e_{z'}I$ in accordance with the phenomenological Eqs.~\eqref{phenom_cubic}, where $\Phi_1 = -\Phi_2$.
Using the energy balance equation
\begin{equation}
	\label{en_balance}
	\sum_{\bm k, \bm k',\pm} W^{(\pm)}_{\bm k'\bm k}(\varepsilon_{k'}-\varepsilon_k)= \alpha I,
\end{equation}
where $\alpha$ is the absorption coefficient, 
we obtain the following estimate 
\begin{equation}
	\label{Phi_1_2}
	\abs{\Phi_{1,2}}
	\approx \xi  {e \tau m g\mu_{\rm B}\over \hbar^3} \alpha.
\end{equation}
Here $m$ is the effective mass for parabolic dispersion, and $m=\varepsilon_{\rm F}/v_0^2$  for a linear dispersion. 

For the intraband absorption considered here, we can rewrite the obtained expression as follows
\begin{equation}
	\label{Phi_1_2_Drude}
	\abs{\Phi_{1,2}}I \approx \xi  {Ne^3  g\mu_{\rm B}\over \hbar^3 \omega^2} \abs{\bm E}^2,
\end{equation}
where $N$ is the electron concentration
Here we took into account that, in the conditions of the experiment $\omega\tau \gg 1$, and 
the absorption coefficient $\alpha \propto 1/\tau$. 
We see that the amplitude of MPGE current
is independent of $\tau$, so
it is determined only by the electron concentration and radiation frequency.

In stronger classical magnetic fields, the asymmetric spin-dependent scattering Eq.~\eqref{U} also results in the MPGE current. Accounting for the Lorentz force in the kinetic equation gives the results~\eqref{phenom_cubic},~\eqref{Phi_1_2} with the CR peak in the absorption coefficient~\cite{Olbrich2013}:
\begin{equation}
	\label{Phi_1_2_CR}
	\abs{\Phi_{1,2}}I \approx \xi  {Ne^3  g\mu_{\rm B}/\hbar^3 \over (\omega\mp \omega_c)^2+1/\tau^2} \abs{\bm E}^2.
\end{equation}
Here 
$\omega_c$ 
is the classical cyclotron frequency for the electrons at the Fermi energy, and the upper and lower sign is for the active ($\bm E \perp \bm B$) and passive ($\bm E \parallel \bm B$) linear polarizations in the Voigt geometry.

We see that the amplitude of MPGE current $j_x \propto B_y\Phi_{1,2}$ has resonant and nonresonant parts. In the active linear polarization, the current has a maximum at CR. 

\subsection{Quantizing magnetic fields}

In higher magnetic fields, when Landau quantization takes place, the 
asymmetrical part of the electron-phonon interaction Eq.~\eqref{U}
also results in the fluxes in the spin subbands and in the MPGE electric current. 
Here we consider 
strong magnetic fields assuming that a complete spin polarization is established. In this case,
according to Eq.~\eqref{j_spin_current} at $S=-1/2$, the MPGE current has a value $\bm j=-2e\bm{\mathcal J}$.

We also assume that 
$\hbar\omega<\Delta$, 
where $\Delta$ is the energy gap between neighboring Landau subbands,
so both the initial and final states of optical transitions lie in the same Landau subband. 
In so strong magnetic fields, a free motion of the carriers is possible along the $\bm B$ direction only. 
Under these conditions, the MPGE current $\bm j \parallel \bm B$ 
is suppressed. This follows from the form of the scattering matrix element in Eq.~\eqref{Upm}, where at $\bm k, \bm k' \parallel \bm B \parallel y$ the term with $\sigma_y$ is absent. 
Analogously, as can be seen from the corresponding Eq.~\eqref{energy_dia}, the orbital mechanism of MPGE also does not give a current $\bm j \parallel \bm B$ for any direction of $\bm B$.

\begin{figure}[h]
	\centering \includegraphics[width=\linewidth]{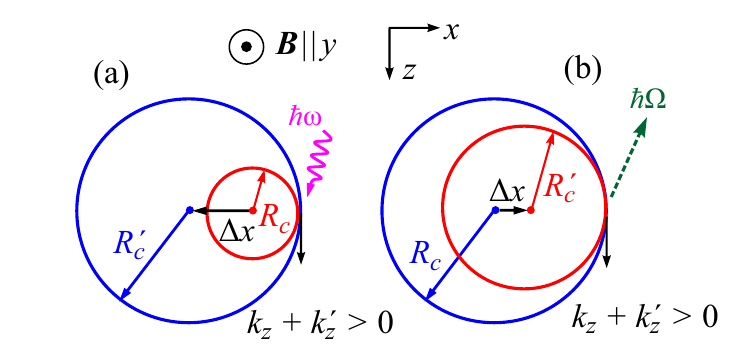} 
	\caption{ Model of the perpendicular to the magnetic field 
		electron flux in one spin subband in the classically strong magnetic field. 
		Circles depict the
		cyclotron orbits in the $(zx)$ plane perpendicular to the magnetic field $\bm B \parallel y$ for the particle with 
		the cyclotron radius $R_c$. Blue and red circles depict
		cyclotron orbits for the carriers with higher and lower energy.
		Allowance for the wavevector-dependent
		contribution in Eq.~\eqref{Upm} makes the electron-phonon scattering with $k_z + k_z' > 0$ more
		probable giving rise to the carrier flow in the given spin-up subband directed along $x$ axis.
		Panel (a): 		
		The shift current formed at nonresonant radiation absorption.
		Red and blue circles depict
		cyclotron orbits for the electron before and after absorption of a photon with energy $\hbar\omega$ accompanied by a scattering on a phonon.
		Due to change of the cyclotron
		radius, the carrier shifts in the real space by $\Delta x = \abs{R_c-R_c'}$, where $R_c'$
		is the cyclotron radius after the absorption process. 
		Panel (b): 		
		The  current formed at cooling of carriers by emission of phonons following a resonant radiation absorption.
		Blue and red circles depict
		cyclotron orbits for the electron before and after emission of a phonon with energy $\hbar\Omega$.
		Due to change of the cyclotron
		radius, the carrier shifts in the real space by $\Delta x = \abs{R_c-R_c'}$, where $R_c'$
		is the cyclotron radius after the phonon emission. 
	}
	\label{Fig_shifts}
\end{figure}

The electric current perpendicular to $\bm B$ is nonzero in quantized magnetic fields.
It is formed due to 
shifts of the center of the cyclotron orbit of photoexcited electrons which occur 
at scattering by phonons~\cite{Zoth2014}, see  Fig.~\ref{Fig_shifts}. The direction of the shifts in the spin subbands is fixed by the odd in momentum terms in the probability of the electron-phonon interaction, Eq.~\eqref{Upm}.

In the nonresonant case, the radiation absorption is accompanied by phonon scattering. Therefore the shift current is formed in the process of absorption, Fig.~\ref{Fig_shifts}(a).
By contrast, 
at resonant absorption, carriers make direct optical transitions between the Landau subbands.
The electric current $\bm j \perp \bm B$ may appear due to a shift of the carrier cyclotron orbit center in the final state in comparison to the initial state of the transition. However, at direct optical transitions such shifts equal to zero. 
This is clear if one takes the Landau gauge  with the vector potential $A_z=-xB_y$, so the the cyclotron orbit center position is $k_zl_B^2$, where $k_z$ is the wavevector of motion in the plane perpendicular to $\bm B$, and $l_B=\sqrt{\hbar/\abs{eB}}$ is the magnetic length.  The matrix element of a direct transition keeps $k_z$ intact, 
therefore the shift of the cyclotron orbit center is absent~\cite{Ivchenko1984}.
The perpendicular to $\bm B$ electric current is formed in the process of energy relaxation which follows the direct optical absorption process~\cite{Zoth2014}.
Fast inter-particle scattering establishes a quasi-equilibrium distribution with a temperature higher than the temperature of the lattice. Then this distribution relaxes via interaction with phonons.
At scattering by phonons, the electron cyclotron orbit center is shifted, Fig.~\ref{Fig_shifts}(b). 

Below we estimate the shift photocurrent $\bm j \perp \bm B$  for nonresonant intraband absorption and resonant CR and R transitions.

\subsubsection{Nonresonant photocurrent}

In the nonresonant case, the spin fluxes in the $x$ direction read~\cite{Zoth2014}
\begin{equation}
	i_{x,s} = \sum_{\bm k, \bm k'}(x_{\bm k} - x_{\bm k'})W_{\bm k' \bm k}^s,
\end{equation}
where $x_{\bm k}=k_zl_B^2$ is cyclotron orbit center, and $W_{\bm k' \bm k}^s$ is the scattering assisted intraband optical absorption probability given by Eq.~\eqref{W}.
This spin flux formation is illustrated in Fig.~\ref{Fig_shifts}(a).
If the Fermi energy exceeds by far the cyclotron energy, then one can ignore an effect of magnetic field on the electron scattering and use the above expression for calculation of the spin fluxes expressing all the values via the quantum numbers in the magnetic field~\cite{Lifshitz1981}. 
Using the energy balance Eq.~\eqref{en_balance},
we obtain the following estimate for the electric current 
\begin{equation}
	\label{jperp}
	j_x=e(i_{x,+}+i_{x,-})\approx e \xi \tilde{k}^2 \sin2\theta\sin\varphi_{\bm B}{\alpha I\over \hbar \omega_c}.
\end{equation}
Here we again assumed a complete spin polarization, and $\tilde{k}$ is a value of the mean wavevector corresponding to 
the Fermi energy or $k_{\rm B}T$ at degenerate statistics and in the Boltzmann gas, respectively. 
Deriving Eq.~\eqref{jperp}, we took into account that 
the coefficients $a_{ijl}$ introduced in Eq.~\eqref{Upm} satisfy the relation $a_{zzz}+a_{zxx}+a_{zyy}=\sin2\theta\sin\varphi_{\bm B}$.

The obtained estimate of $j_x$, Eq.~\eqref{jperp}, is valid for any polarization of the absorbed radiation. The dependence of $j_x$ on the polarization vector $\bm e$ is present in Eq.~\eqref{jperp} in the absorption coefficient $\alpha(\bm e)$.
The nonresonant absorption in a quantized magnetic field is different for polarizations along and perpendicular to $\bm B$.
For $\bm E \parallel \bm B$, the energy absorption rate in the ground Landau subband is given by
a usual high-frequency Drude expression 
\begin{equation}
	\alpha_\parallel I = 2\abs{\bm E}^2{Ne^2 \over m \omega^2 \tau}.
\end{equation}

For the configuration $\bm E \perp \bm B$,  the electron-photon interaction has matrix elements between the neighboring Landau subbands only. Therefore, the  
radiation is absorbed due to virtual transitions via the next Landau subband.
The absorbed power is given by
\begin{equation}
	\alpha_\perp I =\hbar\omega\sum_{k,k'}W_{k' k},
\end{equation}
where $k$, $k'$ are wavevectors in the magnetic field direction, and the probability of the optical transition $k \to k'$ is given by the Fermi Golden rule 
\begin{multline}
	W_{k' k} = {2\pi\over \hbar} \sum_{\nu = \pm} 
	\qty[f_k(1-f_{k'})N_\nu-f_{k'}(1-f_k)N_{-\nu}] 
	\\
	\times 
	\abs{(\tilde{U}_{k' k}+\tilde{U}_{k k'}){V_{n+1,n} \over \Delta}}^2 
	\delta(\varepsilon_{k}+\hbar\omega -\nu\hbar\Omega - \varepsilon_{k'}).
\end{multline}
Here 
$\tilde{U}_{k' k}$ are matrix elements of intersubband scattering by phonons, 
$V_{n+1,n}$ and $\Delta$ are the electron-photon interaction matrix element and the energy gap between the 
Landau subbands with numbers $n$ and $n+1$, 
and other notations are the same as in Eq.~\eqref{W}.

For parabolic energy dispersion we have  
\begin{equation}
	\Delta = \hbar\omega_c, \quad V_{n+1,n} = {ie\over \omega}E {\hbar \sqrt{n+1}\over \sqrt{2}ml_B}.
\end{equation}
As a result,
assuming $\hbar\omega \ll k_{\rm B}T$ and $n=\varepsilon_{\rm F}/\hbar\omega_c \gg 1$, we obtain 
\begin{equation}
	\label{alpha_perp}
	\alpha_\perp I = 4\abs{\bm E}^2{Ne^2 \over m \omega_c^2 \tau}\abs{\tilde{U}\over U}^2,
\end{equation}
where 
the last factor is a squared ratio of the inter- to intrasubband electron-phonon interaction matrix elements.
A similar expression is valid for linear dispersion as well.

It follows from Eqs.~\eqref{jperp} and~\eqref{alpha_perp} that the nonresonant photocurrent $j_x(\bm E \perp \bm B)$ decreases in high magnetic fields according to
\begin{equation}
	\label{nonrez_polariz}
	j_x(\bm E \perp \bm B) \sim {1\over B^3}.
\end{equation}
Since it is linear in $B$ in low fields, Eq.~\eqref{j_perp}, we conclude that it has a maximum as a function of  $B$.

\subsubsection{Resonant  MPGE}
\label{theory_CR}

The shifts of the cyclotron orbit center of resonantly heated carriers occur at their cooling which is carried out via scattering by phonons. 
The direction of the shifts is opposite in two spin subbands and 
dictated by the odd in momentum spin-dependent terms in the probability of the electron-phonon interaction.
It follows from Eq.~\eqref{Upm} that the latter has the form
\begin{equation}
	\mathcal W_{\bm k' \bm k}^\pm= \mathcal W_0\qty(1\pm 2\xi \sum_{i,j,l=x,y,z}a_{ijl}K_iK_jK_l),
\end{equation}
where  $\mathcal W_0$ is the probability of electron scattering by phonons without account for the non-centrosymmetry of the crystal.

The spin fluxes in the $x$ direction read
\begin{equation}
	i_{x,s} = \sum_{\bm k, \bm k'}(x_{\bm k} - x_{\bm k'})\mathcal W_{\bm k' \bm k}^s.
\end{equation}
For carriers' energy relaxation by phonons following the resonant radiation absorption, 
the energy balance equation reads
\begin{equation}
	\sum_{\bm k, \bm k'}(\varepsilon_k-\varepsilon_{k'})\mathcal W_0=\alpha_{\rm res} I,
\end{equation}
where $\alpha_{\rm res}$ is the resonant absorption coefficient. 
Then we obtain the following estimate for the electric current $j_x=e(i_{x,+}+i_{x,-})$
\begin{equation}
	\label{j_res}
	j_x \approx e \xi \tilde{k}^2 \sin2\theta\sin\varphi_{\bm B}{\alpha_{\rm res}(\omega) I\over \hbar \omega_c}.
\end{equation}
Here we again assumed a complete spin polarization, and $\tilde{k}$ 
is defined after Eq.~\eqref{jperp}.
This expression shows that the resonant photocurrent has the same frequency dependence as the absorption coefficient.
It is remarkable that this expression, despite being similar to Eq.~\eqref{jperp} describing the MPGE current formed in the course of radiation absorption, differs strongly because here $\alpha_{\rm res}(\omega)$ is the resonant absorption coefficient of a direct optical transition. 
A difference in microscopics of these two shift current formation is shown in Fig.~\ref{Fig_shifts}.

\section{discussion}
\label{discussion}

\begin{figure*}
	\centering \includegraphics[width=\linewidth]{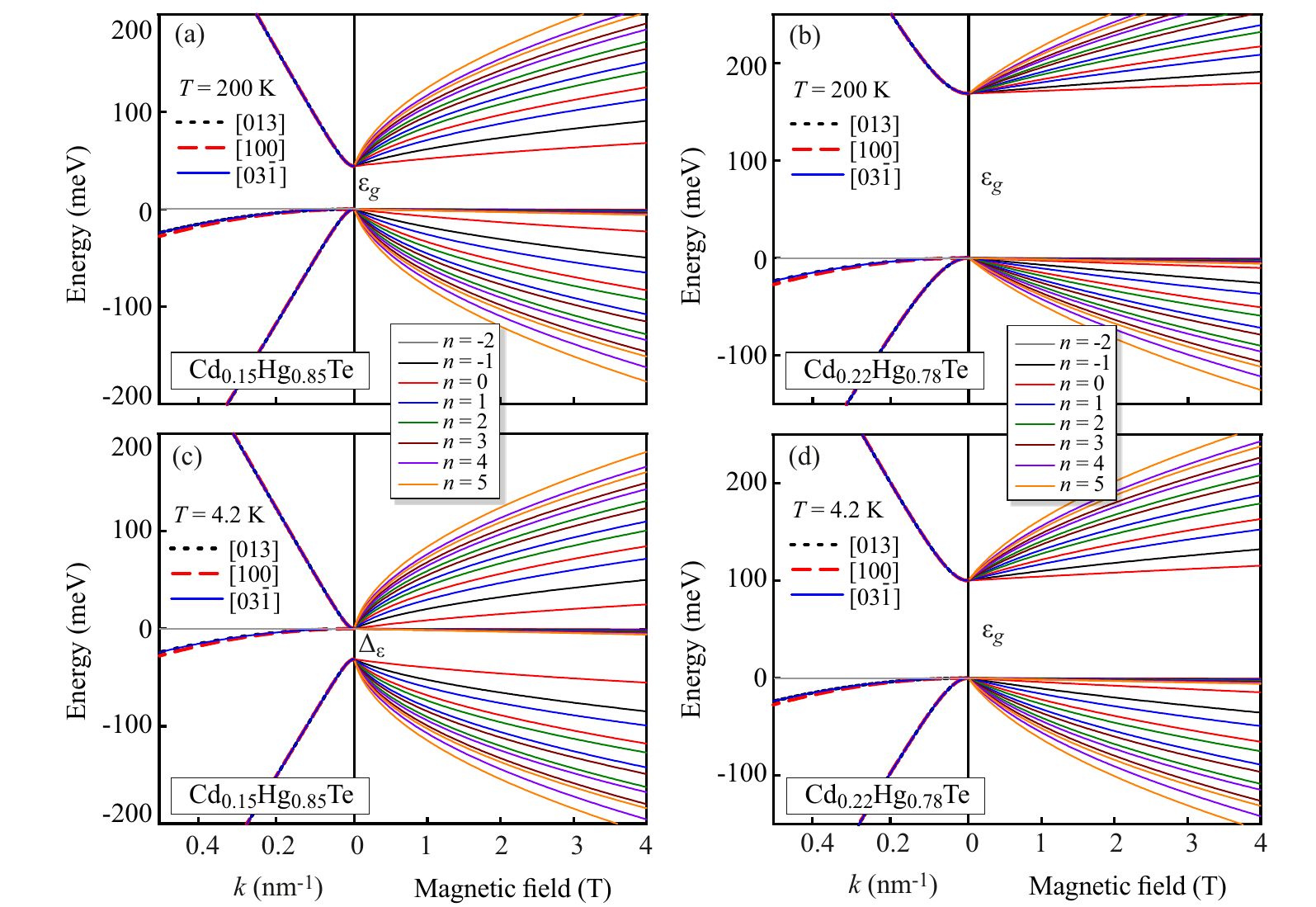}
	\caption{Energy spectra of Cd$_{x}$Hg$_{1-x}$Te with $x=0.15$ and 0.22 calculated for $T=200$ and 4.2~K. Left parts of each panel present the spectra at zero magnetic field.
Here, $\varepsilon_g$ is the band-gap of the bulk film, while $\Delta_{\varepsilon}$ represents the energy gap between $\Gamma_6$ and $\Gamma_8$ bands at the $\Gamma$ point of the Brillouin zone.
Note that Cd$_{0.15}$Hg$_{0.85}$Te has  $\varepsilon_g=0$ at $T=4.2$~K, panel (c).  Right parts show magnetic field dependencies of the Landau levels.
}
\label{Fig13}
\end{figure*}

\begin{figure}
	\centering \includegraphics[width=\linewidth]{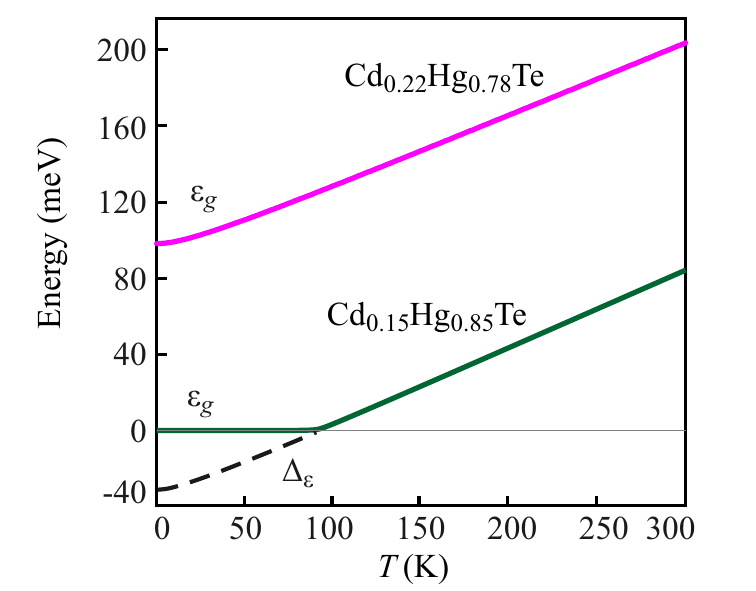}
	\caption{Calculated temperature dependence of the band gap $\varepsilon_g$,
 and the energy gap $\Delta_\varepsilon$ between $\Gamma_6$ and $\Gamma_8$ bands at the $\Gamma$ point of the Brillouin zone in Cd$_{x}$Hg$_{1-x}$Te with $x=0.15$ and 0.22. 
	}
	\label{Fig12}
\end{figure}

\begin{figure*}
\centering \includegraphics[width=\linewidth]{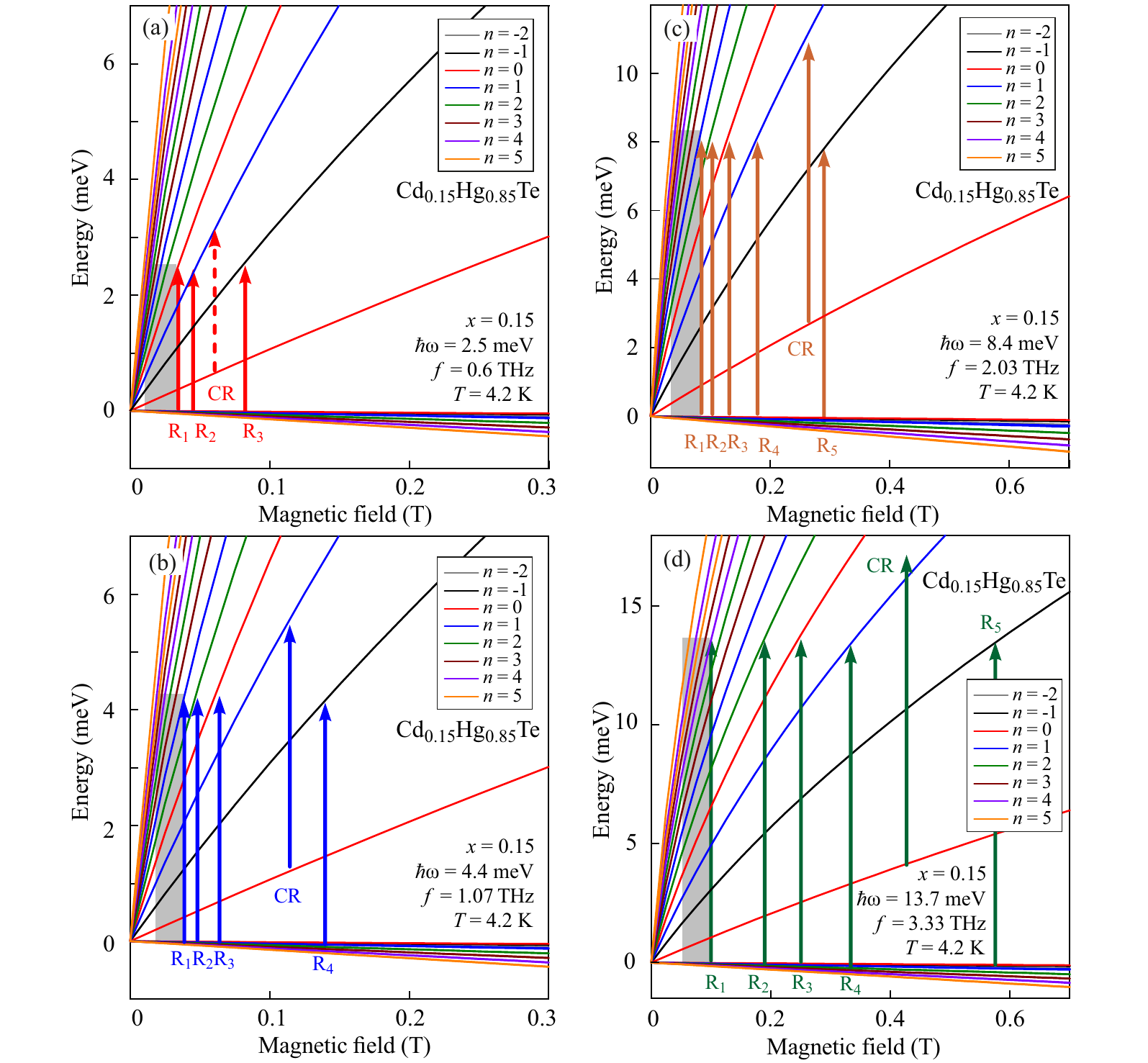}
\caption{ 
	Zoom of the low energy part of magnetic field dependence of the Landau levels  for Cd$_{x}$Hg$_{1-x}$Te with $x=0.15$ calculated for $T=4.2$~K. Vertical arrows show resonant optical transitions detected in experiment, see Fig.~\ref{Fig_JB015}. The arrow are labeled as CR and R$_{1,2,3,4,5}$  which depict  the polarization sensitive and polarization independent resonances, respectively. The length of the arrows corresponds to the photon energies used in our experiments:  $f = 0.6$~THz ($\hbar\omega =2.5$~meV),  $f = 1.07$~THz ($\hbar\omega =4.4$~meV), 2.03~THz ($\hbar\omega =8.4$~meV), and 3.33~THz ($\hbar\omega =13.7$~meV). Gray rectangles sketch possible optical transitions involving high LLs, which can hardly be resolved in our experiments. For the magnetic fields that correspond to the sketched resonances, see Table~\ref{table1}.
}
\label{Fig_LLs015_wide_ab}
\end{figure*}

\subsection{Nonresonant MPGE}

The theory presented above explains our experimental findings described in Sec.~\ref{Experiment}, including the origin of the nonresonant and resonant MPGE currents, the photocurrent behavior upon variation of the magnetic field, as well as its polarization and frequency dependencies.

At high temperatures, the resonant MPGE is not detected and the photoresponse is caused by the non-resonant MPGE in the whole range of magnetic fields.  At low $B$, the photocurrent has opposite directions for $\pm B$ and increases linearly with increasing magnetic field, see Figs.~\ref{Fig2} and \ref{Fig10}, and radiation intensity, as it is expected from Eqs.~\eqref{phenom_cubic},~\eqref{j_perp}. The observed polarization dependence, see Figs.	\ref{Fig3} and \ref{Fig11}, is also in line with the theory. In the experiments, the angle $\alpha$ defining the polarization state was reckoned from the axis in the sample plane which is perpendicular to $\bm B$, Fig.~\ref{Fig14}. 
In two  samples under study, the angle between $\bm B$ and the [100] axis is $\varphi_{\bm B} \approx \pm 45^\circ$, see Fig.~\ref{Fig14}. Therefore we have from Eq.~\eqref{j_perp} 
\begin{equation}
	\label{j_perp_alpha}
	j_\perp = \pm {{B}\cos{2\theta}\over 2} \qty(\Phi_1\cos{2\alpha}  - \Phi_2)I,
\end{equation}
where the upper and lower signs are for samples A and B, respectively, where signals by chance were picked up along two perpendicular crystallographic directions. We see that in both cases the current $j_\perp$ contains the polarization independent part and the part depending on the radiation polarization as $\propto \cos2\alpha$, as detected in experiment, see Figs.	\ref{Fig3} and \ref{Fig11}.

Figures~\ref{Fig2} and~\ref{Fig3} demonstrate that the photocurrent excited by radiation with lower frequency ($f=0.6$~THz) is about 4 times smaller than that in response to the radiation with $f= 1.07$~THz. This difference follows from Eq.~\eqref{Phi_1_2_Drude} which yields for the frequency dependence of the MPGE current $j\propto \omega^{-2}$.

At low temperatures and low magnetic field the nonresonant MPGE is hidden by a much stronger resonant MPGE, see Figs.~\ref{Fig4}, \ref{Fig5}, \ref{Fig_JB015}, \ref{Fig8}, and \ref{Fig9}. However, it is clearly detected at high magnetic fields and for some frequencies shows up as raising nonresonant photoresponse at low magnetic fields, too. In contrast to the low magnetic field regime, at high magnetic field it decreases upon the magnetic field increase, which is also in agreement with the theory. It follows from Eqs.~\eqref{j_perp} and~\eqref{nonrez_polariz} that $j_x(\bm E \perp \bm B)$ raises linearly at low fields and drops as $1/B^3$ at high fields. This allows one to approximate the whole magnetic-field dependence of the nonresonant part by an empirical formula
\begin{equation}
	\label{empiric}
j_x={aB\over cB^4+d}.
\end{equation}
The corresponding fits are presented in Figs. \ref{Fig4}. The decrease of the signal at high fields is also detected at high temperatures, however, it is not only decreases with raising $B$, but even changes its sign, see Figs.~\ref{Fig2} and \ref{Fig10}. We attribute this fact to the interplay of the spin and orbital mechanisms of the MPGE, see Eqs.~\eqref{U} and~\eqref{energy_dia}, which give opposite signs of the photocurrent, and approach maximum at different magnetic field values.

\subsection{Resonant MPGE}

\subsubsection{CR induced MPGE in sample with $x = 0.15$}

We begin with the discussion of the resonant MPGE detected in sample A with $x=0.15$, see Figs.~\ref{Fig4}, \ref{Fig5}, and~\ref{Fig_JB015}. Natural candidates for the resonant MPGE are the photocurrents caused by the cyclotron resonance (CR) and inter Landau level transitions. In the Voigt geometry used in our experiments the CR can only be excited if the radiation  electric field is oriented normally to the magnetic field, $\bm E(\omega) \perp \bm B$. These conditions fulfill the resonances labeled as CR in Figs.~\ref{Fig5} and \ref{Fig_JB015}. 
To confirm this conclusion and to identify multiple R-type resonances we calculated the energy dispersion of Cd$_{0.15}$Hg$_{0.85}$Te and analyzed possible optical transitions for all frequencies used in our experiments, see Figs. \ref{Fig13},~\ref{Fig12}, and~\ref{Fig_LLs015_wide_ab}.

\begin{table}[]
		\caption{Resonance positions and the cyclotron mass $m_{\rm CR}$ obtained for sample A.
}
			\begin{tabular}{|ccccccccc|}
		\hline
		\multicolumn{9}{|c|}{Data for Cd$_{0.15}$Hg$_{0.85}$Te}                                                                                                                                                                                                                                         \\ \hline
		\multicolumn{1}{|c}{} & \multicolumn{1}{c|}{} & \multicolumn{6}{c|}{Resonance position, mT}                                                                                                                               & \\ \hline
		\multicolumn{1}{|c|}{$\hbar\omega$, meV}                   & \multicolumn{1}{c|}{$\lambda$, $\mu$m}                  & \multicolumn{1}{c||}{CR}  & \multicolumn{1}{c|}{$R_1$} & \multicolumn{1}{c|}{$R_2$} & \multicolumn{1}{c|}{$R_3$} & \multicolumn{1}{c|}{$R_4$} & \multicolumn{1}{c||}{$R_5$} &     CR mass, $m_{\rm CR}/m_0$                          \\ \hline
		\multicolumn{1}{|c|}{13.7}               & \multicolumn{1}{c|}{90}                & \multicolumn{1}{c||}{430} & \multicolumn{1}{c|}{100}   & \multicolumn{1}{c|}{190}   & \multicolumn{1}{c|}{250}   & \multicolumn{1}{c|}{330}   & \multicolumn{1}{c||}{560}   & 0.0036                          \\ \hline
		\multicolumn{1}{|c|}{8.4}                & \multicolumn{1}{c|}{148}               & \multicolumn{1}{c||}{265} & \multicolumn{1}{c|}{85}    & \multicolumn{1}{c|}{100}   & \multicolumn{1}{c|}{133}   & \multicolumn{1}{c|}{180}   & \multicolumn{1}{c||}{290}   & 0.0036                          \\ \hline
		\multicolumn{1}{|c|}{4.4}                & \multicolumn{1}{c|}{280}               & \multicolumn{1}{c||}{115} & \multicolumn{1}{c|}{39}    & \multicolumn{1}{c|}{48}    & \multicolumn{1}{c|}{64}    & \multicolumn{1}{c|}{140}   & \multicolumn{1}{c||}{--}      & 0.0031                          \\ \hline
		\multicolumn{1}{|c|}{2.5}                & \multicolumn{1}{c|}{496}               & \multicolumn{1}{c||}{60}  & \multicolumn{1}{c|}{33}    & \multicolumn{1}{c|}{45}    & \multicolumn{1}{c|}{82}    & \multicolumn{1}{c|}{--}      & \multicolumn{1}{c||}{--}      & 0.0028                          \\ \hline
	\end{tabular}
			\label{table1}
\end{table}

\begin{table}[h]
	\caption{Resonance positions and the cyclotron mass $m_{\rm CR}$ obtained for sample B.
}
	\begin{tabular}{|ccccc|}
		\hline
		\multicolumn{5}{|c|}{Data for Cd$_{0.22}$Hg$_{0.78}$Te}                                                                                                                             \\ \hline
		\multicolumn{2}{|c|}{}                                                            & \multicolumn{2}{c|}{Resonance position, mT}                   &                                 \\ \hline
		\multicolumn{1}{|c|}{$\hbar\omega$, meV} & \multicolumn{1}{c|}{$\lambda$, $\mu$m} & \multicolumn{1}{c|}{CR}   & \multicolumn{1}{c|}{$\mathcal R$} & CR mass, $m_{\rm CR}/m_0$ \\ \hline
		\multicolumn{1}{|c|}{13.7}               & \multicolumn{1}{c|}{90}                & \multicolumn{1}{c|}{1140} & \multicolumn{1}{c|}{--}             & 0.0095                          \\ \hline
		\multicolumn{1}{|c|}{8.4}                & \multicolumn{1}{c|}{148}               & \multicolumn{1}{c|}{630}  & \multicolumn{1}{c|}{125}         & 0.0086                          \\ \hline
		\multicolumn{1}{|c|}{4.4}                & \multicolumn{1}{c|}{280}               & \multicolumn{1}{c|}{300}  & \multicolumn{1}{c|}{290}          & 0.0078                          \\ \hline
		\multicolumn{1}{|c|}{2.5}                & \multicolumn{1}{c|}{496}               & \multicolumn{1}{c|}{145}  & \multicolumn{1}{c|}{620}          & 0.0067                          \\ \hline
	\end{tabular}
	\label{table2}
\end{table}

Figures  \ref{Fig13} and \ref{Fig_LLs015_wide_ab} show the band structure and Landau levels (LLs) calculated using the eight-band $\bm k\cdot \bm p$ Kane Hamiltonian, which directly takes into account the interactions between $\Gamma_6$, $\Gamma_8$, and $\Gamma_7$ bands 
(neglecting the contribution due to the bulk inversion asymmetry).
To calculate the energy of LLs, we apply the axial approximation by omitting the warping terms in the Hamiltonian. In this case, the electron-wave function for a given LL index $n > 0$ generally has eight components, describing the contribution of the $\Gamma_6$, $\Gamma_8$, and $\Gamma_7$ bands into the LL. We note that a specific LL with $n=-2$ contains only a contribution of the heavy-hole band with a momentum projection $\pm 3/2$. Details of the LL notation, explicit form of the Hamiltonian and temperature dependent band-structure parameters are provided in Ref.~\cite{Krishtopenko2016}.

The observed resonances in the MPGE current are described by Eq.~\eqref{j_res}, which shows that the photocurrent is proportional to the absorption coefficient and, therefore,  follows to the absorption Lorentzian, see Figs.~\ref{Fig_JB015}. As next we read out from the experiment plots positions of the polarization dependent resonances labeled as CR and depicted them in Fig.~\ref{Fig_LLs015_wide_ab} by vertical lines with the length corresponding to used  photon energies. Figures~\ref{Fig_LLs015_wide_ab}(b)-(d) show that for radiation frequencies $f = 1.07$~THz ($\hbar\omega =4.4$~meV), 2.03~THz ($\hbar\omega =8.4$~meV), and 3.33~THz ($\hbar\omega =13.7$~meV) the corresponding arrows perfectly fit transitions between the zeroth and the first Kane fermion LLs in the conduction band,
i.e. indeed correspond to the CR. Note that for $f = 0.6$~THz ($\hbar\omega =2.5$~meV) the CR resonance position is very close to the polarization independent R-resonance, which could be the reason why it was not resolved in our experiments, see dashed arrow in Fig.~\ref{Fig_LLs015_wide_ab}(a). 

Electron cyclotron mass and the lowest limit of the momentum relaxation time can be estimated from the resonance position and the full width half maximum (FWHM), see Figs.~\ref{Fig5}, \ref{Fig_LLs015_wide_ab} and the inset in Fig. \ref{Fig9}. We obtained the CR mass increases with raising frequency from 0.032 $m_0$ ($f=1.06$~THz) to 0.0035$m_0$ ($f=3.33$~THz), see Table~\ref{table1}, and that the momentum relaxation time is longer or equal to  0.5~ps. Note that the resonances can be broadened by interference effects or superradiant decay~\cite{Abstreiter1976,Chiu1976,Mikhailov2004,Zhang2014,Herrmann2016}. 
The obtained values confirm that the resonances are caused by the optical transitions in the conduction band and are in agreement with the transport data for electrons, see Appendix~\ref{App_A}. 

We would like to emphasize that, at low temperatures, the magnitude of the resonant and even non-resonant magneto-photogalvanic current $J/P$ is enormously large approaching several  milliAmperes per Watt, see, e.g. Fig.~\ref{Fig4}. To our knowledge it exceeds THz-radiation induced photogalvanic currents in other materials by several orders of magnitude.

\subsubsection{R-type resonant MPGE in sample with $x = 0.15$}

Now we turn to the polarization independent resonances of type R. Figure~\ref{Fig13} shows that the photon energies used in our experiments may result in the interband transitions, because Cd$_{0.15}$Hg$_{0.85}$Te has zero band-gap at $T=4.2$~K, see Fig.~\ref{Fig12}. Note that the energy gap between $\Gamma_6$ and $\Gamma_8$ bands at the $\Gamma$ point is about  40~meV, see Fig.~\ref{Fig13},   that exceeds significantly our photon energies  $\hbar\omega \leq 13.7$~meV. 

\begin{figure}
	\centering \includegraphics[width=\linewidth]{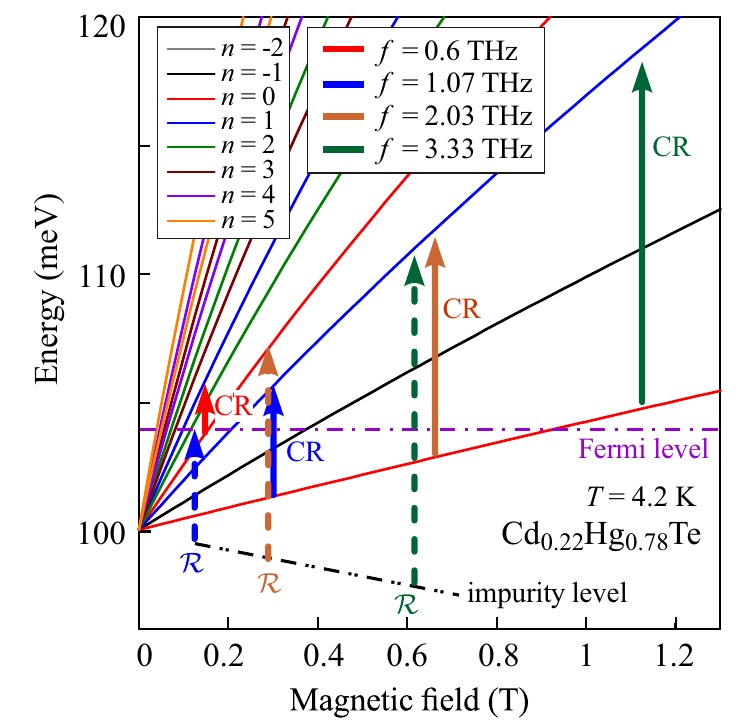}
	\caption{ Zoom of the low energy part of magnetic field dependence of the Landau levels  for Cd$_{x}$Hg$_{1-x}$Te with $x=0.22$ calculated for $T=4.2$~K. Vertical arrows show resonant optical transitions detected in experiment, see Fig.~\ref{Fig9}. For the  corresponding magnetic fields  see Table~\ref{table2}. The arrow are labeled as CR and $\cal{R}$,  which depict  the polarization sensitive and polarization independent resonances, respectively. The length of the arrows corresponds to the photon energies used in our experiments: $f = 0.6$~THz ($\hbar\omega =2.5$~meV);  $f = 1.07$~THz ($\hbar\omega =4.4$~meV); 2.03~THz ($\hbar\omega =8.4$~meV), and 3.33~THz ($\hbar\omega =13.7$~meV).  
	}
	\label{Fig_LLs022}
\end{figure}

Alike for the CR resonances we depicted in Fig.~\ref{Fig_LLs015_wide_ab} all detected R-resonances by vertical arrows with the length corresponding to used  photon energies. One can see that all optical transitions start in the valence band and have final states on one of the electron LLs. We emphasize that those resonances in the terahertz range of frequencies are only possible for Kane fermion energy dispersion resulting in the almost flat valence band and, for $x=0.15$ and low temperatures, in zero band gap. Previously these transitions have been detected in experiments on the radiation transmission~\cite{Ikonnikov2013}. Importantly, according to the selection rules such resonances are allowed in any polarization, see Refs.~\cite{Orlita2014,Yavorskiy2018} and Sec.~\ref{theory_CR}, which completely agrees with experiment. Also in line with experiment the observed resonances in the MPGE current are described by Eq.~\eqref{j_res}, and consequently,  the photocurrent is proportional to the absorption coefficient having a Lorentzian shape, see Fig.~\ref{Fig_JB015}.  Figure~\ref{Fig_JB015} also reveals that the amplitude of the R-resonances are substantially larger than that of the CR one. This observation we attribute to the higher populations of the initial states for R transitions than for CR.
 
To finalize the discussion of the resonances in Cd$_{0.15}$Hg$_{0.85}$Te at $T=4.2$~K we note that, despite under these conditions the film  is characterized by the inverted band structure, we did not find any traces of the MPGE involving Pankratov-Volkov surface states.  This conclusion can be reached because the photocurrent response of two samples with normal and reverse banding does not differ qualitatively.

\subsubsection{CR- and $\mathcal R$-type resonant MPGE in sample with $x = 0.22$}

As next we discuss the resonances detected at $T=4.2$~K in Cd$_{0.22}$Hg$_{0.78}$Te, see Figs.~\ref{Fig8} and \ref{Fig9}, applying the same procedure as we used for the analysis of the resonances in Cd$_{0.15}$Hg$_{0.85}$Te. The results are shown in Fig.~\ref{Fig_LLs022} in which a color code is used to highlight optical transitions excited by the radiation with different photon energies shown by vertical arrows. First of all the polarization dependent resonant MPGE currents in Figs.~\ref{Fig8} and the resonances labeled as CR in Fig.~\ref{Fig9}(a) (red arrow), correspond well to the transitions between the 0th and 1st Kane fermion LLs in the conduction band for all four frequencies. 
The CR electron mass determined from the resonance position monotonic increases with the frequency increase from  $m=0.0067m_0$ ($f=0.6$ THz) to 0.0095$m_0$ ($f=3.33$ THz), see Table~\ref{table2}. The lowest limit of the momentum relaxation time estimated from the FWHM is 1.1~ps.
%
While also in this sample we detected the polarization independent resonances, they can not be attributed to the transitions between the LLs in the valence and conduction bands. As it is seen from Figs.~\ref{Fig13} and~\ref{Fig12}, the band gap in this sample at $T=4.2$~K is about 100~meV being much larger than any of photon energies used in our experiment. In contrast to the multiple  R-resonances in Cd$_{0.15}$Hg$_{0.85}$Te, in Cd$_{0.22}$Hg$_{0.78}$Te we detected only a single $\cal R$-lines for each frequency. Note that the maximum value of the $\cal R$-resonant photocurrent  at the same frequency in the sample B is 4-5 times smaller than R-resonance in sample A. We attributed these resonances to the optical transitions from an impurity level to the first LL, see vertical arrows labeled by $\cal R$ in Fig.~\ref{Fig_LLs022}, which were previously discussed in Ref.~\cite{Littler1990}. The shape of the photocurrent resonances also reflects the one of the absorption due to the direct optical transitions, see Eq.~\eqref{j_res}.

Finally we note, that the samples under study have a lower symmetry than $T_d$ point group of an ideal bulk crystal. This follows from Ref.~\cite{Hubmann2020} where the helicity-dependent photocurrent was detected in the same samples. Microscopically, it is caused by additional compared to Eq.~\eqref{U} terms in the electron-phonon interaction, which may result in the photocurrent contributions forbidden in $T_d$ symmetry, e.g., the circular MPGE, which, however, are out of scope of the present study.

\section{summary}
\label{summary} 

Our studies show that irradiation of Cd$_{x}$Hg$_{1-x}$Te crystals hosting Kane fermions with THz radiation leads to magnetophotogalvanic currents with the magnitude, to our knowledge,  is orders of magnitude that larger than those studied in other systems. By measuring the MPGE current at room and liquid helium temperatures  in crystals with different values of Cd content $x$ we observed  nonresonant and resonant photocurrents. The developed theory, which considers a pure spin current excited by asymmetric spin-dependent scattering and converted into a dc electric current by the Zeeman effect, explains the experimental results well.  The nonresonant MPGE is shown to be caused by Drude absorption assisted by phonon scattering. All the resonances observed at low temperature  in the MPGE current are well described by optical transitions between the adjusting conduction band Landau levels (CR-resonance), interband transitions from the valence band to the higher LL (R-resonances, samples with Cd content $x=0.15$), and optical transitions from the impurity level  of the higher LL levels ($\cal{R}$-resonances, samples with Cd content $x=0.22$). The positions of the CR and multiple R-type resonances  surprisingly well correspond to the  band structure and LL position calculated using the eight-band $\bm k\cdot \bm p$ Kane Hamiltonian, which directly accounts for the interactions between the $\Gamma_6$, $\Gamma_8$, and $\Gamma_7$ bands.

\section{Acknowledgments}
\label{acknow} 
We thank I.~Yachniuk for fruitful discussions. The financial support of the Deutsche Forschungsgemeinschaft (DFG, German Research Foundation) via Project-ID 521083032 (Ga501/19),
the Volkswagen Stiftung Program (97738)
 is gratefully acknowledged.  
We also acknowledge the France 2030 program for Equipex+ HYBAT project (ANR-21-ESRE-0026), the Physics Institute of CNRS for Tremplin 2024 - STEP - project, 
the Occitanie Region for the ``Quantum Technologies Key Challenge'' program - TARFEP - project, and the Terahertz Occitanie Platform.

	\appendix
\counterwithin{figure}{section}
\setcounter{figure}{0}

\section{Analysis of the transport and magnetotransport data obtained for  Cd$_{x}$Hg$_{1-x}$Te with $x=0.15$ and $0.22$.}
\label{App_A}

\begin{figure}
	\centering \includegraphics[width=\linewidth]{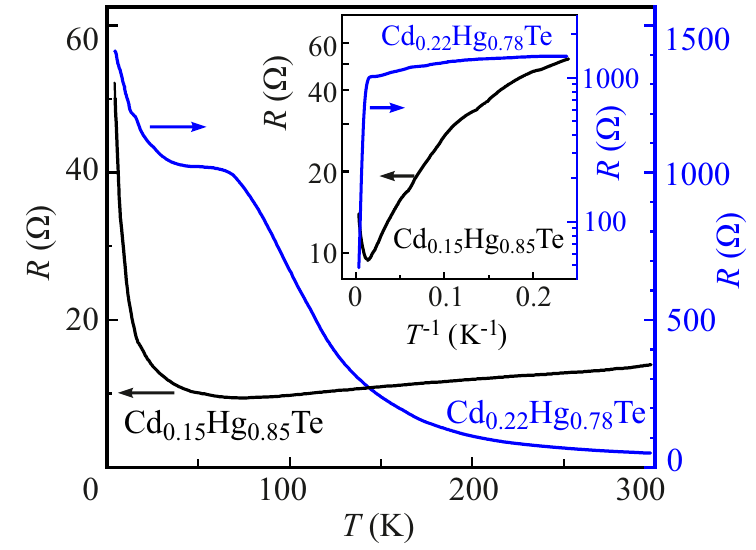}
	\caption{
		Temperature dependence of the two-point resistance measured in samples A (black, left axis) and B (blue, right axis). Inset shows the data re-plotted as Arrhenius plot.  
	}
	\label{figA01}
\end{figure}

The MPGE studies were supplemented by transport measurements performed both on samples A and B and on other samples made of the same substrates (hereinafter referred to as samples $A_1$ 
and $B_1$). The dependences of $R(T)$ in zero magnetic field and at temperatures from 4.2 to 300\,K presented in Fig.~\ref{figA01} show that the resistance of sample $A_1$ varies in the range from 10 to 50\,Ohm, and for sample $B_1$ from 50 to 1500\,Ohm. Despite more than an order of magnitude difference in resistance, the behavior of the samples remains similar. For both investigated samples, the course of the temperature dependence of the resistance is mainly determined by the processes of charge carrier activation as well as phonon scattering. At low temperature, the resistance of both samples saturates at values of 50 and 1500\,Ohm, which is characteristic of metallic or semi-metallic systems. Together with the large value of the band gap at $T=0$ for sample $B_1$ ($\approx$100\,meV, see Fig.~\ref{Fig12}), this indicates a bended zone structure along the $z$-axis, probably arising during the process of film growth, and forming spatially separated electronic and hole layers, degenerated at low temperatures. For sample $A_1$, band bending is also possible, but not necessary, since the effective value of the energy gap, which determines the activation of carriers, is equal to 0 for the inverted spectrum. The resistance of the sample $B_1$ decreases with temperature with two characteristic regions (weakly at low $T$, and then sharply at high $T$, see the inset in Fig.~\ref{figA01}, where the ln$R(1/T)$ dependences are shown), which may be related to the transition from the degenerate to the nondegenerate state. In this case, the crossover temperature (80\,K) should coincide in order of magnitude with the electron Fermi energy, which allows us to estimate it as 7\,meV. By contrast, for sample $A_1$ there is no pronounced activation dependence at any temperature, instead there is a broad minimum of resistance between 50 and 100\,K. This behavior of $R(T)$ reflects the strong dependence of the  band gap $\varepsilon_g$ on temperature and its transition through zero, see Fig.~\ref{Fig12}.

\begin{figure*}
	\includegraphics[width=\linewidth]{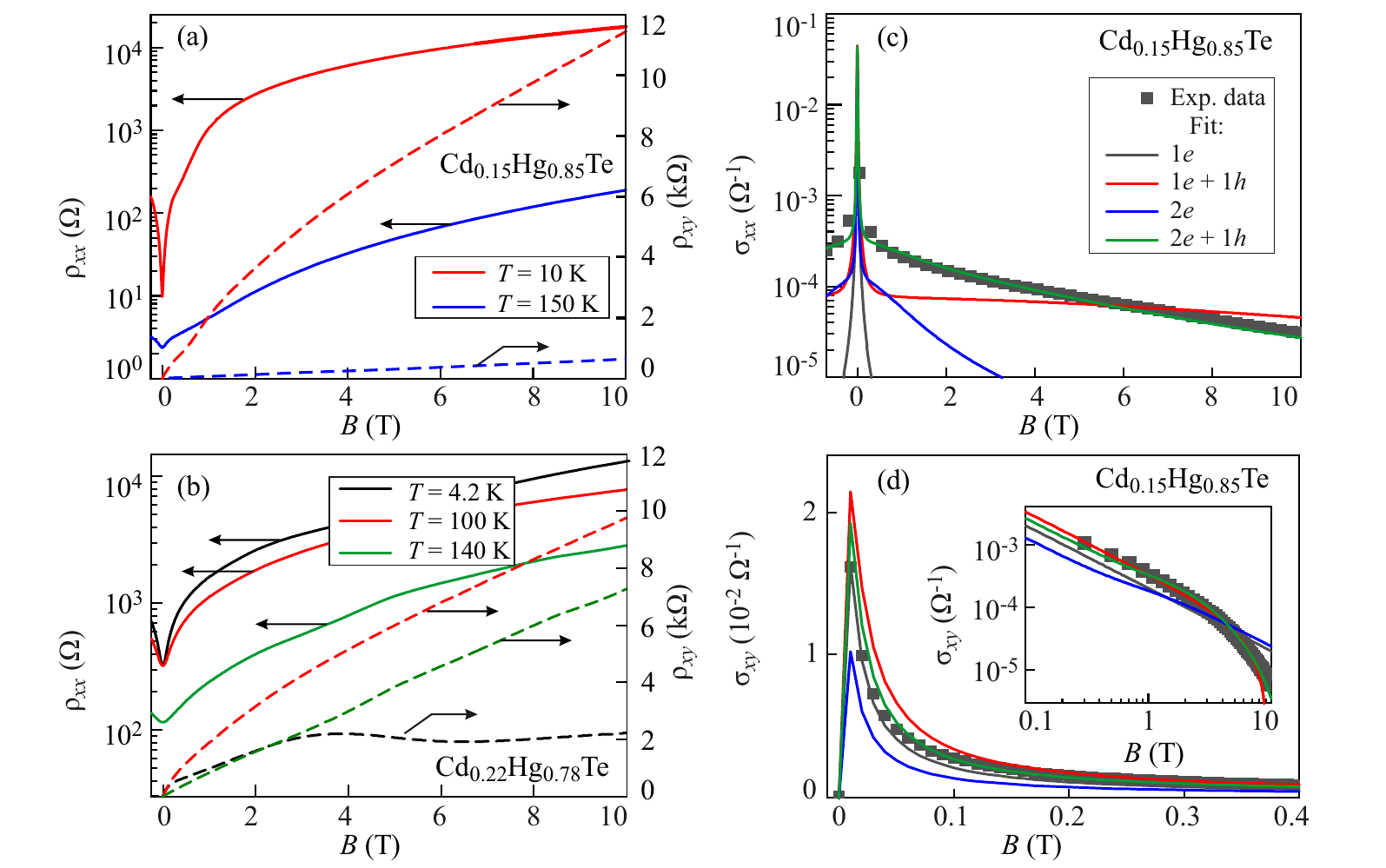}
	\caption{
		(a) and (b): experimental dependences $\rho_{xx}(B)$ (solid lines, left) and $\rho_{xy}(B)$ (dashed, right) for A and B samples, respectively, measured at low and high temperatures. (c) and (d): an example of the fitting of $\sigma_{xx}(B)$ and $\sigma_{xy}(B)$, respectively, by the Drude model with different amount of carrier groups. Experimental data are obtained from the $\rho_{xx}(B)$ and $\rho_{xy}(B)$ dependences for the sample A at a temperature of 1.6\,K and are shown as separate points. The solid lines depict the different fits: one electron type (e, black), one electron and one hole group (e + h, red), two electron groups (2e, blue), and two electron and one hole groups (2e + 1h, green). It is well seen that the last fit follows the experimental data significantly better than all the others. The inset in panel (d) shows the dependences of $\sigma_{xy}(B)$ over a wider range of $B$ and on a logarithmic scale.  
	}
	\label{Drude_Fitting}
\end{figure*}

To identify charge carriers and determine their parameters (density and mobility), magnetotransport measurements were carried out for samples $A_1$ and $B_1$ using a 4-point scheme in the temperature range from 100\,mK to 150\,K and in a magnetic field up to 10\,T, see Fig.~\ref{Drude_Fitting}(a) and (b). The obtained $\rho_{xx}(B)$ and $\rho_{xy}(B)$  dependencies were recalculated into conductivity tensor components $\sigma_{xx}(B)$ and $\sigma_{xy}(B)$ and fitted by the Drude model with multiple carrier types. For simplicity, we consider the system under study as two-dimensional system. This approach is justified in the absence of the current flow along the growth axis of the sample, which is fulfilled in samples of large size (i.e., with negligibly small edge effects) and in the absence of a magnetic field in the sample plane. Because of the significant reduction of $\sigma_{xx}(B)$ and $\sigma_{xy}(B)$ in a strong magnetic field (by almost 4 orders of magnitude relative to the maxima), a weight function was used that proportionally amplifies the contribution of asymptotically small $\sigma_{xx}$  and $\sigma_{xy}$ regions at high $B$. The fitting by several types of carriers was chosen for the following considerations: first, several types of carriers have been observed in thick HgTe quantum wells \cite{Kozlov2014, SAVCHENKO2021114624} and HgTe and CdHgTe films \cite{Savchenko_2023, Yavorskiy2018}, and therefore similar behavior can be expected in the investigated bulk CdHgTe films. In this case, due to the lack of doping and the system's tendency to electroneutrality, the total density of electrons and holes should be comparable; secondly, the simultaneous existence of electrons and holes is indicated by the positive magnetoresistance and strongly nonlinear Hall resistance $\rho_{xy}(B)$, Fig.~\ref{Drude_Fitting}(a) and~(b). Fitting by the simplest model with one type of carriers in this case gives a meaningless result: the average slope of the $\rho_{xy}(B)$ dependence for the sample $B_1$ first increases with increasing temperature, which, using a one-component model, would mean a decrease in the carrier concentration, which contradicts the observed temperature dependence $R(T)$. When using the Drude model with two types of carriers (electrons and holes), the result becomes more plausible, but the addition of a third group of carriers (slow electrons) allows for even better agreement between the fit and experiment. An example of such fits for sample $A_1$ is shown in Fig.~\ref{Drude_Fitting}(c) and~(d).

\begin{figure}[t]
	\centering \includegraphics[width=\linewidth]{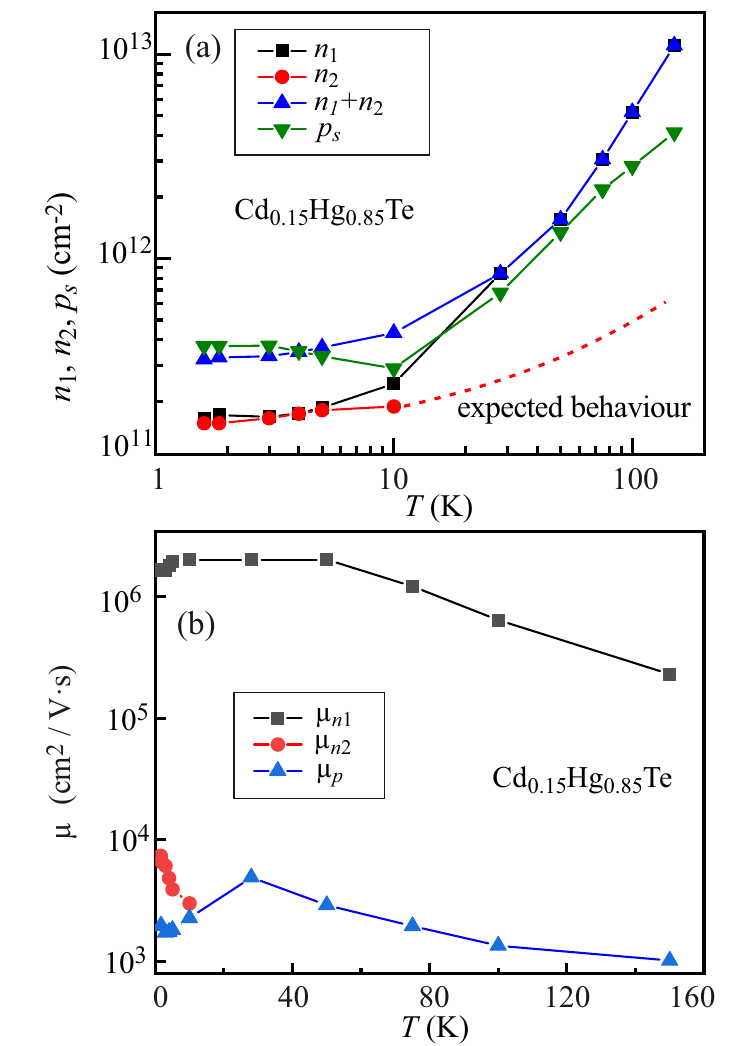}
	\caption{Carrier densities $n$ and $p$ (a) and mobility $\mu$ (b) for the sample $A_1$ extracted from the fit of magneto-transport data shown in Fig.~\ref{Drude_Fitting} applying 3-carriers Drude model. 
		The model 	involves fast electrons ($n_1$), slow electrons ($n_2$) and holes ($p$). Note that at $T>10$\,K the mobility of slow electrons becomes so small, so it makes further reliable identification of these carriers impossible. Thus, the fit for higher $T$ was made only for 2 types of carriers. In this aspect, the increase in mobility $\mu_p$ at $T=25$\,K must be considered as an artifact. The expected density dependence $n_{n2}(T)$ shown by a dashed red line.
	}
	\label{Fitting_Result_A}
\end{figure}

\begin{figure}[t]
	\centering \includegraphics[width=\linewidth]{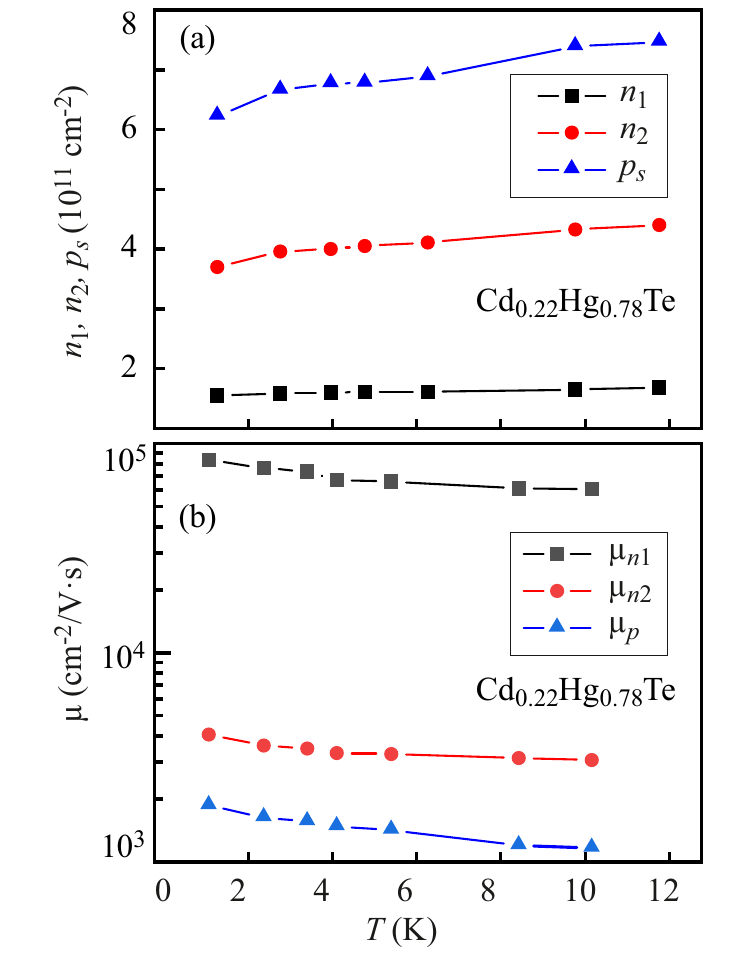}
	\caption{Carrier densities $n$ and $p$ (a) and mobility $\mu$ (b) for the sample $B_1$ extracted from the fit of magneto-transport data shown in Fig.~\ref{Drude_Fitting} applying 3-carriers Drude model. As for type A sample, the model involves fast electrons ($n_1$), slow electrons ($n_2$) and holes ($p$). At $T\geqslant15$\,K the hole mobility becomes smaller than 1000\,cm$^2$/V$\cdot$s, making the whole fitting unreliable. 
	}
	\label{Fitting_Result_B}
\end{figure}

When using the multicomponent Drude model, it is necessary to keep in mind its limitations associated with the impossibility of separating carriers of close mobility and masking of low-mobility carriers due to their small contribution to the conductivity. Another important aspect is the incomplete correspondence between the model, which assumes independent groups of carriers with uniquely defined and independent of the magnetic field parameters, and the real system. Therefore, the use of even three types of carriers is not always justified, and the transition to a continuum distribution (mobility spectrum approach \cite{Yavorskiy2018}) turns out to be a deliberate excess of the model accuracy. The results of the fitting, taking into account the above factors, are presented in Figs.~\ref{Fitting_Result_A} and \ref{Fitting_Result_B} for samples $A_1$ and $B_1$, respectively.

The sample $A_1$ (Fig. \ref{Fitting_Result_A}) at low temperature is characterized by two groups of electrons with equal concentration of about $1.6\times10^{11}$\,cm$^{-2}$, but with different 2 orders of mobility ($10^4$ and $2\times10^6$\,cm$^2$/V$\cdot$s), as well as holes (with concentration $p_{\rm s} = 3.6\times10^{11}$\,cm$^{-2}$ and mobility 2000\,cm$^2$/V$\cdot$s). The resulting densities are two-dimensional as they are naturally obtained by fitting the experimental data with the two-dimensional Drude model. To translate into three-dimensions, the thickness of the layer in which certain carriers are localized is required, the exact value of which remains unknown, thus, we present 2D densities.  The values obtained are in qualitative agreement with recent Ref.~\cite{Yavorskiy2018}, as well as the older work~Ref.~\cite{Fau1984}. With increasing temperature, the mobility of low-mobility electrons decreases rapidly, but the reliability of the fit also decreases. At temperatures above 10\,K, reliable separation of this group of electrons is not possible, and further fitting was performed using only one type of electrons and holes. For the same reason, the maximum of hole mobility is observed in this region, which is an artifact. The high-mobility electrons are apparently bulk carriers (rather than topological surface states) whose high mobility is associated with a low effective mass value. From the point of view of transport measurements, this is confirmed both by the absence of Shubnikov-de Haas oscillations at 80\,mK (not shown) and by the preservation of mobility at high values up to $T=150$\,K, i.e., above the temperature of transition to the normal spectrum.

The type B sample (Fig.~\ref{Fitting_Result_B}) is similarly characterized by two kinds of electrons and holes with densities of the same order ($n_1 = 1.6\times10^{11}$\,cm$^{-2}$, $n_2 = 3.7\times10^{11}$\,cm$^{-2}$ and $p_s = 6.2\times10^{11}$\,cm$^{-2}$). In comparison to the type A sample, here all carriers are characterized by much lower mobility, not exceeding $10^5$\,cm$^2$/V$\cdot$s for fast electrons, 4000\,cm$^2$/V$\cdot$s for slow electrons, and 1900\,cm$^2$/V$\cdot$s for holes. As the temperature increases, the mobility of low-mobility carriers decrease rapidly, which also reduces the reliability of the fit. At $T>15$\,K, the mobility of holes becomes less than 1000\,cm$^2$/V$\cdot$s, which makes further fit too inaccurate.

\bibliography{all_lib1.bib}

\end{document}